\documentclass[aps,twocolumn]{revtex4-2}
\usepackage{amsmath}
\usepackage[colorlinks]{hyperref}
\usepackage[utf8]{inputenc}
\usepackage{bm}
\usepackage{graphicx}

\begin{document}

\title{Effects of dispersion parity-violating interaction in electron scattering and atoms}

\author{V.~V.~Flambaum}
\email{v.flambaum@unsw.edu.au}
\affiliation{School of Physics, University of New South Wales, Sydney 2052, Australia}

\author{I.~B.~Samsonov}
\email{igor.samsonov@unsw.edu.au}
\affiliation{School of Physics, University of New South Wales, Sydney 2052, Australia}

\begin{abstract}
Exchange of two neutrinos (as well as other fermions) generates a long-range parity-violating (PV) potential of the form $\sim G^2/r^5$, with characteristic range $\hbar/(2m_\nu c)$. In atomic systems the corresponding matrix elements converge at distances $r < 10/M_Z$, so that the interaction between electron and quarks effectively reduces to a contact term $\sim G^2 M_Z^2 \delta^{3}(\vec{r}) \sim G\,\alpha\,\delta^{3}(\vec{r})$. This interaction produces a $-0.8\%$ correction to the effective weak charge of cesium, resolving the $2\sigma$ discrepancy between the Standard Model prediction and the measured Cs parity-violation amplitude. The corresponding value of the weak mixing angle is $\sin^2\theta_W = 0.2375(19)$ at $q^2 \approx 0$, in agreement with the Standard Model prediction $\sin^2\theta_W = 0.23873$. The relative correction to the proton weak charge is about $3\%$. Using these results, we revisit the limits on an additional $Z'$ boson and obtain a constraint on isospin-conserving oblique radiative corrections characterized by the Peskin--Takeuchi parameter $S = -0.32(53)$ at $q^2 \approx 0$.
\end{abstract}

\maketitle

\section{Introduction}

\begin{figure*}
    \centering
    \includegraphics[width=0.6\linewidth]{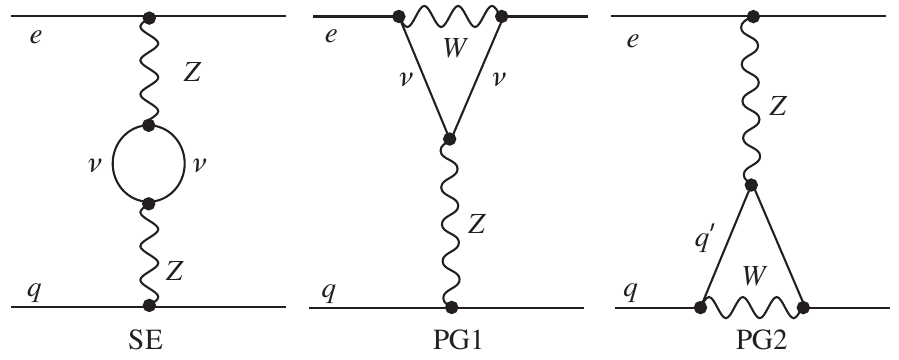}
    \caption{Feynman diagrams contributing to the long-range force due to neutrino exchange between electron $e$ and quark $q$. The self-energy (SE) diagram may also contain other leptons and quarks in the loop while the `penguin' (PG1) diagram has only the electronic neutrino in the loop.}
    \label{fig:diagram1}
\end{figure*}

Exchange of two neutrinos as in Fig.~\ref{fig:diagram1} produces long range potential $\propto G^2/r^5$ in the particle interaction, where $G$ is the Fermi constant \cite{Tamm,Feynman,FeinbergPR1968,Feinberg1989,HsuPRD1994}. The effects of parity conserving part of this potential are, however, many orders of magnitude smaller than the sensitivity of the experiments \cite{Kapner2007,Adelberger2007,Chen2016,Vasilakis2009,Terrano2015,StadnikPRL2018,StadnikErrata,Munro} and are practically unobservable.

In Refs.~\cite{FS07,Ghosh,MunroPV,Ghosh:2024ctv,Ghosh:2025ole} it was noted that the neutrino exchange potential has parity violating (PV) part $\propto G^2/r^5$ with the range $\hbar/2m_{\nu}c$ exceeding atomic size by several orders of magnitude. It was also demonstrated that  mixed $\gamma Z$ electron vacuum polarization  produces PV potential $\propto \alpha G/r^3$ with the  range $\hbar/2m_e c=193$~fm, which exceeds the range of the weak interaction by five orders of magnitude \cite{FS07,MunroPV}, similar to the range of the Uehling  potential due to electron vacuum polarization by the nuclear Coulomb field.

Although the long-distance behavior of the PV potential proportional to $G^2/r^5$ was accurately calculated in Refs.~\cite{FS07,Ghosh,MunroPV}, obtaining a reliable numerical estimate of the effect produced by this potential requires regularization of the singular PV interaction at short distances. In Ref.~\cite{MunroPV}, this was achieved by introducing the $Z$-boson mass into the integral describing the long-range PV potential. While this procedure provided a reasonable numerical estimate of the effect, no actual calculation of the short-distance PV potential was performed. Therefore, the result of Ref.~\cite{MunroPV} should be regarded as an estimate rather than a fully accurate calculation.

The authors of Refs.~\cite{Ghosh:2024ctv,Ghosh:2025ole} showed that the neutrino-exchange potential significantly affects PV matrix elements in muonium, positronium, and atoms. In particular, the PV matrix element in Cs atom changes by $\sim-0.3\%$, improving the agreement of the Standard Model prediction with experiments \cite{Wood}. In the present Letter, we extend this analysis by including the contributions of other fermions in the loop.

The contribution of the diagrams in  Fig.~\ref{fig:diagram1} with exchange by two neutrinos and  by other fermions to the PV  effects in atoms is $\sim$1\%, and this significantly exceeds the error 0.35\% of the PNC measurement in Cs atom \cite{Wood} and the error $<$0.5\% in the many-body atomic  calculations of the $Z$-boson contribution \cite{DzuFlaSus89a,DzuFlaGin02,Blundell90,Blundell92,Ginges05,PBD09,PBD10,CsPNC12}. The work is in progress to improve both experimental~\cite{Elliott1,Elliott2,Elliott3} and theoretical~\cite{TXD22} accuracy. 

The error in atomic calculations cancels out in  the ratio of the PV amplitudes in different isotopes of the same atom~\cite{DzuFlaKhr86,BDF09,VF19}. The measurement for the chain of isotopes of Yb atom was performed in Ref.~\cite{Budker19}. 

The study of PV in atoms provides high precision tests of the Standard Model and new physics beyond it \cite{RevPNC,RevPNC1}.

We will also show that the diagrams on Fig.~\ref{fig:diagram1} give significant contributions to the PV effects in low-energy scattering of electron from proton, deuteron and other nuclei. The corresponding corrections to the proton and nuclear  weak charges are found. A significant correction to PV effects in electron-electron scattering and to the electron weak charge is also expected, but it will be studied elsewhere.

Weak charge $Q_W$ is the constant of the electron-nucleon weak interaction due to the $Z$-boson exchange which has interaction range $r_Z=\hbar/M_Z c\approx0.002$ fm. On the nuclear and atomic scales this may be considered as a Fermi-type contact interaction. 

Radiative corrections to the proton and neutron weak charges of the order $\alpha \approx 1/137$ have been calculated in Refs.~\cite{Marciano,MarcianoSanda,MarcianoSirlin1984,Thomas,Duch:2026ujd} and were shown to exceed 1\%.  However, diagrams on  Fig.~\ref{fig:diagram1}, which actually give comparable contributions,  have not been included into these calculations. The point is that at large distance these diagrams give interaction  proportional to  $G^2$, which is very small. However, this interaction is highly singular, $\propto G^2/r^5$, and integration over distance with electron wave functions and  effective cut-off at $r \sim 1/M_Z$ gives contribution $\sim G^2 M_Z^2 \sim  \alpha G$. Thus, the suppression relative to the ordinary weak interaction mediated by the $Z$-boson exchange is $\sim \alpha$ \cite{MunroPV}. 

For electron-electron scattering there are also box diagrams with crossing electron and crossing neutrinos lines. Note that in the case of Majorana neutrinos, there is an additional box diagram with neutrinos as the vertical sides. Potentially, this could give a possibility to identify Majorana neutrinos using PV in electron scattering, but the contribution of this diagram is small.        

To perform accurate calculations, we require the effective potential generated by the diagrams shown in Fig.~\ref{fig:diagram1} at distances $r\sim 1/M_Z$. The neutrino-exchange contribution to this potential, valid for all distances satisfying $r\ll 1/m_\nu$, was calculated in Refs.~\cite{Ghosh:2024ctv,Ghosh:2025ole}, and our analysis is based on these results. We further include the additional contributions arising from other fermions in the loop.

In this letter, we use natural units with $\hbar=c=1$.

\section{Effective potential due to the fermion loop}
Parity violating interactions produce PV effects in atoms by mixing $s_{1/2}$ and $p_{1/2}$ electron orbitals. To find corrections produced by diagrams on Fig.~\ref{fig:diagram1}  it is sufficient to calculate the ratio of the matrix elements 
\begin{equation}
\frac{\langle ns_{1/2}|W_f(r)|n'p_{1/2}\rangle}{\langle ns_{1/2}|W_Z(r)|n'p_{1/2}\rangle}\,,
\label{e:LQ}
\end{equation}
where $W_Z$ and $W_f$ are the PV potentials produced by $Z$ boson tree-level exchange and $f$ fermion loop correspondingly.

Dominating contribution to both matrix elements comes from the distance between electron and quark of order $r \sim 1/M_Z$, which is very small as compared to the scale of the variation of electron wave functions and nuclear size. Therefore, we may replace both potentials  by their  Fermi-type contact approximations $w_{Z,f} \delta^3(\vec r)$ using relation $\int W_{Z,f}(r) 4 \pi r^2 dr= w_{Z,f} \int \delta^3(\vec r) d^3r$. 
Then the  nuclear-spin-independent (NSI) part of the electron-nucleus  PV operator $W(r)$ may be cast in the form 
\begin{equation}
 W(r)=-\frac{G}{2\sqrt{2}} (Q_W + \delta Q_W) \gamma_5  \rho(r)\,, \label{e:Qw}\,
\end{equation}
where $Q_W = -N+ (1- 4 \sin^2\theta_W) Z$ is the weak nuclear charge,  $\delta Q_W$ is the correction produced by the diagrams on Fig.~\ref{fig:diagram1},  $\rho(r)$ is the nuclear density normalized by the condition $\int \rho(r) d^3r=1$, $N$ and $Z$ are numbers of neutrons and protons in the nucleus, respectively. Inclusion of radiative corrections gives \cite{PDG2024RPP,PDG2024SM}
\begin{align}
Q_W =& -0.98897N +(1.0250- 3.9978 \sin^2\theta_W)Z 
\nonumber\\
=& -0.98897N +0.070605 Z
\,.
\label{e:Qwx}
\end{align}
For cesium atom, this formula implies the standard model value of the weak charge $Q_W^{SM}=-73.26(1)$ assuming $\sin^2\theta_W=0.23873$ at zero momentum transfer \cite{PDG2024SM}. The experimental measurements of the parity violation in cesium atom give $\sin^2\theta_W=0.2349(18)$ at $q=2.4$ MeV \cite{PDG2024SM}.

Note that the differences in the experimental values of Cs weak charge and Weinberg angle
in 2024 and 2022 editions of Particle Data group reviews are due to the different values of the Stark polarizability $\beta$ used (Cs PV experiment measures ratio of the PV amplitude to $\beta$). The 2023 paper \cite{BaoBeta} corrected the result of Ref.~\cite{Vasil} and confirmed the original value of $\beta$ found in our paper \cite{DzubaBeta}.    

Calculations of the neutrino  exchange potentials, presented in Appendix \ref{AppA}, are similar to that described in detail in Ref.~\cite{Ghosh:2024ctv}. Let us start from the self-energy (SE) diagram in Fig.~\ref{fig:diagram1}. The interaction constant on the bottom fermion line cancels with the single $Z$-boson exchange contribution in the ratio (\ref{e:LQ}). Therefore, for any fermion on the bottom line (e.g. quark or electron in the low energy electron-electron scattering case) the ratio of the SE diagram to the tree-level $Z$-boson contribution is the same. As a result we obtain
\begin{equation}
 \frac{\delta Q_W^{(\text{SE})}}{Q_W} =-\frac{g^2}{96 \pi^2  c_W^2 } N_\text{eff} = -0.86\%\,, \label{e:Qw1} 
\end{equation}
where  $g\approx0.65$ is the $Z$-boson-neutrino interaction constant, $c_W^2=1-s^2_W=0.76127$, $s_W^2\equiv \sin^2\theta_W$.
Following Refs.~\cite{StadnikPRL2018,MunroPV},  we introduced here effective number of neutrinos,
\begin{equation}
N_\text{eff}\approx 14.5\,.
\label{Neff}
\end{equation}
Indeed, there are 3 ordinary neutrinos. However, we should add one-loop contributions of all fermions with masses smaller than that of the $Z$-boson, including $\, e,\,\mu,\, \tau, \, u,\, d,\, s,\, c,\, b$.   $N_\text{eff}$ is not an integer number since interaction constants of other fermions with $Z$-boson are different from that for neutrinos, see Appendix \ref{AppB} for details.

Diagrams PG1 and PG2 in Fig.~\ref{fig:diagram1} do not have enhancement factor $N_\text{eff}$. Using the calculations of the momentum and radial integrals presented in Appendix \ref{AppA}, we obtain the following correction to the nuclear weak charge:
\begin{equation}
 \frac{\delta Q_W^{(\text{PG})}}{Q_W} =1.7
 \times 10^{-3}\frac{N-Z}{N- (1-4 s^2_W ) Z}  \,, \label{e:Qw23} 
\end{equation}
For the proton weak charge this ratio, enhanced by the small value of $(1-4 \sin^2\theta_W )=0.044$ in the denominator,  is 3.8\%. Adding contribution (\ref{e:Qw1}) corresponding to the SE diagram, we obtain the total correction to the nuclear weak charge    
\begin{equation}
 \frac{\delta Q_W}{Q_W}  = \left(-0.86 +0.17 \frac{N-Z}{N- (1-4 s^2_W )Z}\right)\% \,. \label{e:Qw123} 
\end{equation}

Equation (\ref{e:Qw123}) shows that the contribution of the diagrams PG1 and PG2 (given in the last term) for all atoms, except for hydrogen, is small, so we obtain $\delta Q_W/Q_W \approx -0.8\%$ in all medium and heavy atoms.  For the proton weak charge this gives,
\begin{equation}
 \frac{\delta q_W^{p}}{q_W^{p}}
 \approx  3.0\% \,. \label{e:Qwp} 
\end{equation}
Similar relative correction to the PV effect in muonium atom was obtained in Refs.~\cite{Ghosh:2024ctv,Ghosh:2025ole}, where the authors do not interpret this as a correction to the weak charge.  

The enhancement due to the small value of $(1-4s_W^2)$ in the denominator is also expected for the electron weak charge in the low-energy parity-violating electron-electron scattering. Therefore, the relative correction to the electron weak charge may exceed $1\%$.

Distribution of proton and neutron densities in the nucleus are slightly different. Therefore to take into account this effect in numerical calculations it may be convenient to separate proton and neutron contributions to the weak charge. Adding the corrections (\ref{e:Qw123}) to Eq.~(\ref{e:Qwx}) we find
\begin{align}
 Q_W+\delta Q_W  =& -0.98207 N + (1.0181 -3.9634 s^2_W) Z \nonumber\\
 =&-0.98207 N + 0.071918 Z \,.
\label{e:Qwd}
\end{align}
Here we assume the SM value of the Weinberg angle with $\sin^2\theta_W=0.23873$ at zero momentum transfer \cite{PDG2024SM}.

Recall that the aim of the present paper is to calculate the contribution of the PV potential to the effective weak charge in Eq.~(\ref{e:Qw}) which includes the nuclear density function $\rho(r)$. This function, however, drops out from the ratio of the PV potential contribution to the standard $Z$-boson contribution $\delta Q_W/Q_W$. Calculations of atomic matrix elements of the potential $W(r)$ in  Eq.~(\ref{e:Qw}) performed in previous works \cite{DzuFlaSus89a,DzuFlaGin02,Blundell90,Blundell92,Ginges05,PBD09,PBD10,CsPNC12} do take into account of $\rho(r)$, which is usually described by a Fermi-type distribution fitted to measured proton and neutron densities. Understanding that the effects of the nuclear density distribution have been fully addressed in these papers, we do not need to consider them here.

\section{Comparison with experimental data on weak charges and limits on new physics}

If we use the value of the proton weak charge obtained in  the recent very accurate lattice calculation \cite{Zhang:2026utt}, $Q_W^{SM}=0.06987(50)$, we obtain corrected value $Q_W^{SM} + \delta Q_W^{SM} =0.06987 (1+ 0.03)= 0.0719$. It  agrees with  $Q_W^\text{exp}= 0.0719(45)$ measured by Qweak collaboration \cite{Qweak}.

The correction to the electron weak charge is also expected  at the level of several per cent. This does not contradict to the experimental value of the electron weak charge $Q_W^\text{exp}= -0.053(11)$ obtained by E158 collaboration \cite{E158}. The accuracy in the experimental measurement of the proton weak charge in the electron-proton scattering by the Qweak collaboration  is 7\%, the accuracy in the measurement of the PV asymmetry in electron-electron scattering by E158 collaboration is 20\%. However, both experimental collaborations are expecting significant improvements of the accuracy soon.

The most sensitive point to the weak charge correction is currently in cesium PV experiments, where $\delta Q_W/Q_W \approx -0.8\%$ exceeds both the experimental error 0.35\% and the error in  our atomic many-body calculation which is under 0.5\%. This correction actually eliminates $2\sigma$ difference between the measured $Q_W^\text{exp}= -72.41(42)$ and the standard model $Q_W^{SM} + \delta Q_W^{SM} =-73.26 (1-0.008)=-72.67$ weak charges.  

After including $\delta Q_W^{SM}$, the measured in Cs PV experiment value of the Weinberg angle at $q^2\approx 0$, 
\begin{equation}
\sin^2\theta_W=0.2375(19)\,,
\end{equation}
also comes to a perfect agreement with prediction of the standard  model $\sin^2\theta_W=0.2387$. 

The difference $Q_W^\text{exp} - (Q_W^{SM} + \delta Q_W^{SM}) = 0.26(42)$ allows us to obtain limits on new physics. The limits on the additional  $Z'$-boson  interaction strength obtained as a function of $Z'$ mass from the measurements of PV in Cs atom are approximately 2 times stronger than that presented in Table II and Fig.~1 of Ref.~\cite{Dzuba2017PRL}. For $m_{Z'}\gg \alpha Zm_e$, the products of $Z'$ interaction constants with the electron and nucleus obeys the following bound: $|g_e^A g_N^V|/m_{Z'}^2<2.2\times 10^{-8}$ GeV$^{-2}$. For $m_{Z'}\ll 1/ R_\text{atom}$, the product of these constants is limited by $|g_e^A g_N^V|<1.7\times 10^{-14}$.

We also obtain limits on oblique radiative corrections produced by new particles.  Ref.~\cite{MarcianoSTU} calculated the shift of the weak charge produced by these corrections  $\Delta Q_W=-0.8 S$,  where $S$ is the Peskin--Takeuchi parameter characterizing isospin-conserving vacuum-polarization effects. Comparing this expression with the measured value $\Delta Q_W=0.26(42)$, we obtain
\begin{equation}
S=-0.32(53)\,.
\end{equation}
The global electroweak fit of $S$ \cite{PDG2024SM} is dominated by the $Z$-pole and $W$-mass measurements, while atomic PV probes oblique radiative corrections at very low momentum transfer, $q^2\approx0$, thereby testing their running over a wide range of scales. Light weakly interacting states can induce momentum-dependent vacuum-polarization effects that contribute differently at $q^2\approx0$ and $q^2\sim m_Z^2$. In this case, PV in cesium provides sensitivity to new physics effects that may not be fully captured by high-energy fits.

\emph{Note added.} After completion of this work and presenting our results in arXiv, Gorchtein and Spiesberger~\cite{Gorchtein:2026ctl} argued that the two-neutrino-exchange contribution to atomic parity violation in ordinary atoms is eliminated, for practical purposes, by finite nuclear-size effects. We disagree with using the coherent nuclear elastic form factor as an ultraviolet cutoff on the electroweak loop before the short-distance matching is performed. The contribution considered in the present work is a radiative correction to the pointlike Standard-Model electron--fermion amplitude, matched at virtual momenta of order $M_Z$ onto local parity-violating electron--quark operators. Nuclear structure enters only subsequently, through matrix elements of these local quark currents, equivalently through the proton and neutron density distributions in the atomic problem. Multiplying the loop amplitude by the nuclear form factor from the outset (effectively using nuclear radius as a short distance cut-off) treats the nucleus as a smooth elementary source at distances at which the loop resolves its constituents, and therefore removes the short-distance Wilson coefficient generated by pointlike electron--quark interactions. In our treatment finite nuclear size smears the resulting contact interaction over the nuclear density, but does not replace the short-distance scale by the nuclear radius. This is the origin of the difference between the conclusion of Ref.~\cite{Gorchtein:2026ctl} and the result obtained here.

Thus the approach of Ref.~\cite{Gorchtein:2026ctl} describes the coherent elastic electron--nucleus component of the finite-size potential, whereas the correction discussed here is the short-distance radiative correction to the local weak charge.

\vspace{3mm}
\textit{Acknowledgments.} -- The work was supported by the Australian Research Council Grant No.\ DP230101058.

\appendix

\section{Reduction of effective interaction potentials to the contact interaction}
\label{AppA}

The exchange of various SM particles other than photon between two point-like fermions  may be described by potential forces which may be considered as a perturbation to the Coulomb interaction potential. If such a potential $V(r)$ is short-range as compared with the size of the nucleus, they may be effectively reduced to the contact interaction with the the delta-function potential,
\begin{equation}
    V_\text{contact}(\vec r) = \kappa \delta^{3}(\vec r)\,,
    \label{Vcontact}
\end{equation}
with some coefficient $\kappa$ defined by
\begin{equation}
    \kappa = \int V(r) d^3r\,.
\end{equation}

The canonical example is the exchange of the $Z$ boson which is described by the potential
\begin{equation}
    V_Z(r) = - \frac{g^2}{64\pi c_W^2} \frac{e^{-m_Z r}}{r}\,,
\end{equation}
where $g\approx 0.65$ is the gauge coupling constant of the $SU(2)_L$ group, and $c_W\equiv \cos\theta_W \approx 0.87$ is the cos of Weinberg angle. We omit here Dirac matrix  $\gamma_5$. Since the range of this interaction is $r=O( m_Z^{-1})$, in atomic calculations it may be replaced with the contact interaction (\ref{Vcontact}) with the coefficient
\begin{equation}
    \kappa_Z = - \frac{g^2}{16 c_W^2 m_Z^2}\,.
    \label{kappaZ}
\end{equation}

Below, we consider effective potentials corresponding to the exchange of neutrinos as per the Feynman graphs in Fig.~1. In the present work, we need two such potentials, which are usually referred to as the self-energy (SE) and `penguin' (PG) contributions, respectively. These potentials were found in the works \cite{Ghosh:2025ole,Ghosh:2024ctv}. We will calculate the corresponding contact interaction coefficients $\kappa$ for these potentials.

\subsection{Neutrino self-energy diagram}

The potential corresponding to the massless neutrino one-loop self-energy diagram was calculated in Refs.~\cite{Ghosh:2025ole,Ghosh:2024ctv}:
\begin{equation}
  V_\text{SE}(r) = -V_{0,\text{SE}} \frac1r \int_0^\infty dt \frac{te^{-r\sqrt{t}}}{(t-m_Z^2)^2}\,,
  \label{V-SE}
\end{equation}
where
\begin{equation}
  V_{0,\text{SE}} = \frac{g^4}{(4c_W)^4}\frac1{24\pi^3}\,.
\end{equation}
The integral in Eq.~(\ref{V-SE}) is divergent because of the double pole in the denominator. This double pole is regularized by a small shift along the imaginary axis, and by retaining the principal value of the integral \cite{Ghosh:2024ctv}. As a result, one finds
\begin{align}
  V_\text{SE}(r) =& \frac{V_0}{2r}\Big[
      e^{m_Z r}(2+m_Z r)\mathrm{Ei}(-m_Z r)
      \nonumber\\&
      + e^{-m_Z r}(2-m_Z r)\mathrm{Ei}(m_Z r) + 2
    \Big],
\label{VSE}
\end{align}
where
\begin{equation}
    \mathrm{Ei}(x) \equiv -\int_{-x}^\infty t^{-1}e^{-t}dt\,.
\end{equation}

The asymptotic behavior of the potential (\ref{VSE}) is
\begin{equation}
  V(r)|_{r\gg m_Z^{-1}} = -\frac{g^4}{2\pi^3 c_W^4 r^5}\,.
\end{equation}
This potential plays significant role at the distance $r\sim m_Z^{-1}$, and decays rapidly at $r\gg m_Z^{-1}$. Hence, in atomic calculations, this potential may be replaced with the contact interaction potential (\ref{Vcontact}) where the coefficient $\kappa$ is found to be
\begin{equation}
  \kappa_\text{SE} = 4\pi\int_0^\infty r^2 V(r)dr = \frac{4\pi V_{0,\text{SE}}}{m_Z^2}\,.
  \label{Vnorm}
\end{equation}
Here the following identity was used:
\begin{equation}
  \int_0^\infty \left[
      e^x(2+x)\mathrm{Ei}(-x) + e^{-x}(2-x)\mathrm{Ei}(x) + 2
    \right]xdx =2\,.
\end{equation}
Comparing Eq.~(\ref{Vnorm}) with Eq.~(\ref{kappaZ}) we find
\begin{equation}
    \frac{\kappa_\text{SE}}{\kappa_Z} = - \frac{g^2}{96\pi^2 c_W^2} \approx - 5.9\times 10^{-4}\,.
\end{equation}


\subsection{Triangle PG diagram contribution to the potential}
\label{AppA2}

The contribution to the potential from the triangle PG diagrams was found in Refs.~\cite{Ghosh:2025ole,Ghosh:2024ctv} in the following form:
\begin{subequations}
\label{VPG-all}
\begin{align}
    V_\text{PG}(r) =& -V_{0,\text{PG}} \frac1r \text{Re} I(r)\,,\label{PGpotential}\\
    I(r)=&\int_0^\infty dt
    \,e^{-\sqrt{t}r}
    \nonumber\\&\times
    \frac{t(2m_W^2 + 3t) - 2(m_W^2 + t)^2 \ln\left(1+\frac{t}{m_W^2} \right)}{t^2(t-m_Z^2 + i0)}\,,
    \label{B2}
\end{align}
\end{subequations}
where
\begin{equation}
    V_{0,\text{PG}} = \frac{g^4}{2048\pi^3 c_W^2}\,.
\end{equation}
Here the principal value of the integral is assumed near the pole $t=m_Z^2$. In Ref.~\cite{Ghosh:2024ctv}, it was shown that the potential (\ref{VPG-all}) possesses the following asymptotic behavior 
\begin{equation}
    V_\text{PG}(r)|_{r\gg m_Z^{-1}} = -\frac{g^4}{256\pi^3 c_W^2 m_Z^2 m_W^2 r^5}\,.
\end{equation}
Thus, this potential drops rapidly and its effects become negligible at the distance of a few $m_Z^{-1}$. In atomic calculations, within the current accuracy goal the potential corresponding to the PG diagram may be replaced by the contact interaction (\ref{Vcontact}) with the corresponding coefficient $\kappa$.

To compare the potential (\ref{PGpotential}) with contact interaction (\ref{Vcontact}), we need to calculate the coefficient
\begin{equation}
    \kappa_\text{PG} = -4\pi V_{0,\text{PG}} \int_0^\infty I(r)r\,dr\,.
    \label{B4}
\end{equation}
Substituting here Eq.~(\ref{B2}) and integrating over $r$, we find
\begin{align}
    I_0\equiv & \int_0^\infty I(r)r\,dr 
    \nonumber\\ = & \frac1{m_W^2}
    \int_0^\infty \frac{x(2+3x) - 2(1+x)^2\ln(1+x)}{x^3(x-z)}\,.
\end{align}
where $z\equiv \frac{m_z^2}{m_W^2}$. Making use of the identity
\begin{equation}
    \ln(1+x) = x^2 - \frac{x^2}{2} + x^3\int_0^1 \frac{\alpha^2d\alpha}{1+\alpha x}\,,
\end{equation}
reduces the problem to the integration of rational functions,
\begin{align}
    I_0 = &\frac1{m_W^2} \int_0^\infty \frac{xdx}{x-z}\nonumber\\
    &- \frac2{m_W^2}\int_0^1 \alpha^2 d\alpha \int_0^\infty \frac{(1+x)^2dx}{(1+\alpha x)(x-z)}\,.
\end{align}
It is possible to show that all divergent terms in these integrals cancel, and the finite result is 
\begin{align}
    I_0 =& \frac1{2z^3 m_W^2}\Big[ 4 (z+1)^2 \text{Li}_2(-z)
    +4 (z+1)^2 \ln (z) \ln (z+1)
    \nonumber\\&
    +z (7 z-2 (3 z+2) \ln (z)+4)\Big].
\end{align}
Recalling that $z = c_W^{-2}$, we find
\begin{align}
    I_0 =& \frac1{2m_Z^2}\Big[ 4 \left(c_W^2+1\right)^2 \text{Li}_2\left(-c_W^{-2}\right)
    +4 c_W^2
    \nonumber\\&
    -2 \log
   \left(c_W\right) \big(-4 c_W^2
   +4 \left(c_W^2+1\right)^2 \log
   \left(c_W^{-2}+1\right)
   \nonumber\\&
   -6\big)+7 \Big]
   \approx -\frac{0.92}{m_Z^2}\,.
\end{align}
As a result, the coefficient (\ref{B4}) is found to be:
\begin{equation}
    \kappa_\text{PG} \approx 15 V_{0,\text{PG}}m_Z^{-2}\approx 5.4\times 10^{-5}m_Z^{-2}\,.
\end{equation}
Comparing this coefficient with Eq.~(\ref{kappaZ}), we find
\begin{equation}
    \frac{\kappa_\text{PG}}{\kappa_Z} \approx - 1.6\times 10^{-3}\,.
    \label{EqA23}
\end{equation}


\section{Calculation of vertex constant prefactor}
\label{AppB}

In this paper, we study long-range forces between electron and nucleon with the range $r\gg O(m_Z^{-1})$. This interaction is mediated by SM particles with masses $m \ll m_Z$, which are $\nu^\alpha = (\nu_e,\nu_\mu,\nu_\tau)$ for neutrinos, $l^\alpha =(e,\mu,\tau)$ for electron, muon and tau, $q_k=(u,d,s,c,b)$ for quarks. The relevant part of the SM Lagrangian is
\begin{align}
    {\cal L} = & \frac{g}{4c_W} [
    \bar\nu^\alpha \gamma^\mu (1-\gamma_5)\nu^\alpha
    +\bar l^\alpha \gamma^\mu (c_V^\alpha - c_A^\alpha \gamma_5 )l^\alpha
    \nonumber\\&
    +\bar q_k\gamma^\mu (c_V^{q_k} - c_A^{q_k} \gamma_5) q_k ]Z_\mu \nonumber\\&+
    \frac{g}{2\sqrt2} [ \bar l^\alpha \gamma^\mu(1-\gamma_5) \nu^\alpha W^-_\mu 
    \nonumber\\&
    + \bar u \gamma^\mu (1 - \gamma^5)d  W^+_\mu + h.c. ]\,.
\end{align}
Here we ignore the the CKM matrix because the quarks in the loop may be effectively considered as massless.

The values of the vector and axial-vector constants are given in Table~\ref{Tab:constants}.
\begin{table}
\begin{center}
\begin{tabular}{c|c|c}
         & $c_V$ & $c_A$ \\\hline
$l^\alpha$ & $4s_W^2-1$ & $-1$ \\
$u,c,t$       & $1-\frac83s_W^2$ & 1\\ 
$d,s,b$        & $\frac43 s_W^2 -1 $ & $-1$\\
$p$ & $1-4s_W^2$ & $1$ \\ 
$n$ & $-1$ & $-1$
\end{tabular}
\end{center}
\caption{Values of the vector $c_V$ and axial-vector $c_A$ interaction constants for leptons $l^\alpha = (e,\mu,\tau)$, quarks, proton $p$ and neutron $n$. $s_W^2 = \sin^2\theta_W$, where $\theta_W$ is the Weinberg angle.}
\label{Tab:constants}
\end{table}

\subsection{Electron-proton effective interaction}
\label{AppB1}

At high momentum transfer, the proton is represented by three valence quarks $p=uud$. In the tree-level diagram, $Z$ boson interacts with each of these quarks, thus, $c_V^p = 2c_V^u + c_V^d = 1- 4s_W^2$. Hence, the tree-level $Z$-boson exchange diagram contains the following combination of wave functions
\begin{align}
    &[\bar e \gamma^\mu(c_V^e - c_A^e\gamma_5)e][\bar p \gamma_\mu(c_V^\mu - c_A^\mu\gamma_5)p]
    \nonumber\\
    =&-c_A^e c_V^p [\bar e \gamma^\mu \gamma_5 e][\bar p \gamma_\mu p] +\ldots\,, 
\end{align}
and the corresponding prefactor is
\begin{equation}
    -c_A^e c_V^p = 1-4s_W^2\,.
    \label{eq9}
\end{equation}
The same prefactor appears in the SE diagram in Fig~1. Therefore, these vertex constants cancel out in the ratio of atomic matrix elements.

The SE diagram is, however, enhanced by the factor $N_\text{eff}$. In addition to the three neutrinos, this factor receives the following contributions. There are thee left leptons with the $Z$-boson interaction constant $\frac12(c_V^l + c_A^l) = 2s_W^2-1$, and three right leptons with vertex constants $\frac12(c_V^l - c_A^l) = 2s_W^2$. Then there are $2\times 3$ ($u$ and $c$ quarks with three colors) left up-like quarks with the $Z$-boson interaction constants $\frac12(c_V^u + c_A^u) = 1 - \frac43 s_W^2$, and the same number of the corresponding right quarks with the vertex constants $\frac12(c_V^u - c_A^u) = - \frac43 s_W^2$. Finally, there are $3\times 3$  ($d$, $s$, $b$ with three colors) right down-like quarks with the vertex constant $\frac12(c_V^d + c_A^d) = \frac23 s_W^2 -1$, and the same number of the corresponding right quarks with the constant $\frac12(c_V^d - c_A^d) = \frac23 s_W^2$. All these coefficients enter the SE diagram in Fig.~1 quadratically. Combining them we find
\begin{equation}
    N_\text{eff} = \frac{160}{3}s_W^4 - 40s_W^2 +21\approx 14.5\,.
\end{equation}

The PG1 diagram in Fig.~1 has loop on the electron line, and $Z$-boson interacts with the proton line which consists of the three valence quarks. Therefore, this diagram contains the following combination of the wave functions (see also Appendix A in Ref.~\cite{Ghosh:2024ctv}):
\begin{equation}
    [\bar e\gamma^\mu(1-\gamma_5)e][\bar p \gamma_\mu(c_V^p - c_A^p\gamma_5)p] = -c_V^p[\bar e\gamma^\mu\gamma_5e][\bar p \gamma_\mu p] + \ldots
\end{equation}

There are three PG2 diagrams with the loop on the quark lines: two diagrams with $u\to d+W^+$, and one diagram with $d\to u + W^-$. These diagrams contain the following combinations of wave functions:
\begin{align}
    [\bar e\gamma^\mu(-\gamma_5c_A^e)e][\bar d \gamma_\mu c_V^u d] =& -c_A^e c_V^u [\bar e\gamma^\mu\gamma_5e][\bar d \gamma_\mu d] \\
    [\bar e\gamma^\mu(-\gamma_5c_A^e)e][\bar u \gamma_\mu c_V^d u] =& -c_A^e c_V^d [\bar e\gamma^\mu\gamma_5e][\bar u \gamma_\mu u]\,.
\end{align}
The combined prefactor is
\begin{equation}
    -c_V^p -2c_A^e c_V^d - c_A^e c_V^u = 4s_W^2 -1 -1=-2 + 4s_W^2\,.
    \label{eq13}
\end{equation}
The ratio of prefactors (\ref{eq13}) and (\ref{eq9}) is
\begin{equation}
    -\frac{2-4s_W^2}{1-4s_W^2} \approx -23\,.
    \label{B9}
\end{equation}
The corresponding relative correction to the proton weak charge is the product of Eqs.~(\ref{EqA23}) and (\ref{B9}):
\begin{equation}
    \frac{\delta Q_p}{Q_p} = -\frac{2-4s_W^2}{1-4s_W^2} \frac{\kappa_\text{PG}}{\kappa_Z} \approx \frac{1.7\times 10^{-3}}{1-4s_W^2}\,.
    \label{B10}
\end{equation}

\subsection{Electron-neutron effective interaction}

Neutron is represented by the three valence quarks, $n = udd$. The corresponding $Z$-boson vector coupling is $c_V^n = c_V^u + 2 c_V^d = -1$. Therefore, tree-level $Z$-boson exchange diagram has the prefactor 
\begin{equation}
    -c_A^e c_V^n = -1\,.
\label{eq16}
\end{equation}
The same prefactor appears in the SE diagram, which is enhanced by the effective number of neutrinos $N_\text{eff}$.

In the PG1 diagram, there is the following combination of wave functions:
\begin{equation}
    [\bar e\gamma^\mu(1-\gamma_5)e][\bar n \gamma_\mu(c_V^n - c_A^n\gamma_5)n] = -c_V^n[\bar e\gamma^\mu\gamma_5e][\bar n \gamma_\mu n] + \ldots
\end{equation}
There are also three PG2 diagrams: one diagram with $u\to d + W^+$, and two with $d\to u+W^-$. These diagrams contain the following combinations of wave functions:
\begin{align}
    [\bar e\gamma^\mu(-\gamma_5c_A^e)e][\bar u \gamma_\mu c_V^d u] =& -c_A^e c_V^d[\bar e\gamma^\mu\gamma_5e][\bar u \gamma_\mu u] \\
    [\bar e\gamma^\mu(-\gamma_5c_A^e)e][\bar d \gamma_\mu c_V^u d] =& -c_A^e c_V^u[\bar e\gamma^\mu\gamma_5e][\bar d \gamma_\mu d].
\end{align}
The combined prefactor is
\begin{equation}
    -c_V^n -c_A^e c_V^d - 2c_A^e c_V^u = 2-4s_W^2 \,.
\label{eq20}
\end{equation}
The ratio of prefactors (\ref{eq20}) and (\ref{eq16}) is
\begin{equation}
    \frac{2-4s_W^2}{-1} = -(2-4s_W^2)\,.
    \label{B16}
\end{equation}
The corresponding relative correction to the neutron weak charge is the product of Eqs.~(\ref{EqA23}) and (\ref{B16}):
\begin{equation}
    \frac{\delta Q_n}{Q_n} = -(2-4s_W^2)\frac{\kappa_\text{PG}}{\kappa_Z} \approx 1.7\times 10^{-3}\,.
    \label{B17}
\end{equation}

\subsection{Contribution to the nuclear weak charge due to PG diagrams}
\label{AppB3}

Combination of Eqs.~(\ref{B10}) and (\ref{B17}) gives us the correction to the weak charge of a nucleus composed of $Z$ protons and $N$ neutrons due to the PG diagrams,
\begin{equation}
    \delta Q^{(\text{PG})} \approx \frac{1.7\times 10^{-3}}{1-4s_W^2}Z\cdot Q_p + 1.7\times 10^{-3} N\cdot Q_n\,,
\end{equation}
and substituting here tree-level weak charges $Q_p = 1-4s_W^2$ and $Q_n = -1$, we have
\begin{equation}
    \delta Q^{(\text{PG})} \approx 1.7\times 10^{-3}(Z-N)\,.
\end{equation}
Dividing this expression by the tree-level value of the nuclear weak charge, $Q_W = -N + (1-4s_W^2)Z$, we obtain the contribution
\begin{equation}
 \frac{\delta Q_W^{(\text{PG})}}{Q_W} =1.7
 \times 10^{-3}\frac{N-Z}{N- (1-4 s^2_W ) Z}  \,. 
\end{equation}


%


\begin{thebibliography}{54}%
\makeatletter
\providecommand \@ifxundefined [1]{%
 \@ifx{#1\undefined}
}%
\providecommand \@ifnum [1]{%
 \ifnum #1\expandafter \@firstoftwo
 \else \expandafter \@secondoftwo
 \fi
}%
\providecommand \@ifx [1]{%
 \ifx #1\expandafter \@firstoftwo
 \else \expandafter \@secondoftwo
 \fi
}%
\providecommand \natexlab [1]{#1}%
\providecommand \enquote  [1]{``#1''}%
\providecommand \bibnamefont  [1]{#1}%
\providecommand \bibfnamefont [1]{#1}%
\providecommand \citenamefont [1]{#1}%
\providecommand \href@noop [0]{\@secondoftwo}%
\providecommand \href [0]{\begingroup \@sanitize@url \@href}%
\providecommand \@href[1]{\@@startlink{#1}\@@href}%
\providecommand \@@href[1]{\endgroup#1\@@endlink}%
\providecommand \@sanitize@url [0]{\catcode `\\12\catcode `\$12\catcode
  `\&12\catcode `\#12\catcode `\^12\catcode `\_12\catcode `\%12\relax}%
\providecommand \@@startlink[1]{}%
\providecommand \@@endlink[0]{}%
\providecommand \url  [0]{\begingroup\@sanitize@url \@url }%
\providecommand \@url [1]{\endgroup\@href {#1}{\urlprefix }}%
\providecommand \urlprefix  [0]{URL }%
\providecommand \Eprint [0]{\href }%
\providecommand \doibase [0]{https://doi.org/}%
\providecommand \selectlanguage [0]{\@gobble}%
\providecommand \bibinfo  [0]{\@secondoftwo}%
\providecommand \bibfield  [0]{\@secondoftwo}%
\providecommand \translation [1]{[#1]}%
\providecommand \BibitemOpen [0]{}%
\providecommand \bibitemStop [0]{}%
\providecommand \bibitemNoStop [0]{.\EOS\space}%
\providecommand \EOS [0]{\spacefactor3000\relax}%
\providecommand \BibitemShut  [1]{\csname bibitem#1\endcsname}%
\let\auto@bib@innerbib\@empty
\bibitem [{\citenamefont {{Tamm I.}}(1934)}]{Tamm}%
  \BibitemOpen
  \bibfield  {author} {\bibinfo {author} {\bibnamefont {{Tamm I.}}},\
  }\bibfield  {title} {\bibinfo {title} {{Interaction of neutrons and
  protons}},\ }\href {https://doi.org/10.1038/1341010c0} {\bibfield  {journal}
  {\bibinfo  {journal} {Nature}\ }\textbf {\bibinfo {volume} {134}},\ \bibinfo
  {pages} {1010} (\bibinfo {year} {1934})}\BibitemShut {NoStop}%
\bibitem [{\citenamefont {Feynman}(1996)}]{Feynman}%
  \BibitemOpen
  \bibfield  {author} {\bibinfo {author} {\bibfnamefont {R.~P.}\ \bibnamefont
  {Feynman}},\ }\href@noop {} {\emph {\bibinfo {title} {Feynman Lectures on
  Gravitation}}},\ edited by\ \bibinfo {editor} {\bibfnamefont {F.~B.}\
  \bibnamefont {Morinigo}}, \bibinfo {editor} {\bibfnamefont {W.~G.}\
  \bibnamefont {Wagner}},\ and\ \bibinfo {editor} {\bibfnamefont
  {B.}~\bibnamefont {Hatfield}}\ (\bibinfo  {publisher} {Addison-Wesley},\
  \bibinfo {address} {Reading, MA},\ \bibinfo {year} {1996})\BibitemShut
  {NoStop}%
\bibitem [{\citenamefont {Feinberg}\ and\ \citenamefont
  {Sucher}(1968)}]{FeinbergPR1968}%
  \BibitemOpen
  \bibfield  {author} {\bibinfo {author} {\bibfnamefont {G.}~\bibnamefont
  {Feinberg}}\ and\ \bibinfo {author} {\bibfnamefont {J.}~\bibnamefont
  {Sucher}},\ }\bibfield  {title} {\bibinfo {title} {Long-range forces from
  neutrino-pair exchange},\ }\href {https://doi.org/10.1103/PhysRev.166.1638}
  {\bibfield  {journal} {\bibinfo  {journal} {Phys. Rev.}\ }\textbf {\bibinfo
  {volume} {166}},\ \bibinfo {pages} {1638} (\bibinfo {year}
  {1968})}\BibitemShut {NoStop}%
\bibitem [{\citenamefont {Feinberg}\ \emph {et~al.}(1989)\citenamefont
  {Feinberg}, \citenamefont {Sucher},\ and\ \citenamefont {Au}}]{Feinberg1989}%
  \BibitemOpen
  \bibfield  {author} {\bibinfo {author} {\bibfnamefont {G.}~\bibnamefont
  {Feinberg}}, \bibinfo {author} {\bibfnamefont {J.}~\bibnamefont {Sucher}},\
  and\ \bibinfo {author} {\bibfnamefont {C.-K.}\ \bibnamefont {Au}},\
  }\bibfield  {title} {\bibinfo {title} {The dispersion theory of dispersion
  forces},\ }\href
  {https://doi.org/https://doi.org/10.1016/0370-1573(89)90111-7} {\bibfield
  {journal} {\bibinfo  {journal} {Phys. Rep.}\ }\textbf {\bibinfo {volume}
  {180}},\ \bibinfo {pages} {83} (\bibinfo {year} {1989})}\BibitemShut
  {NoStop}%
\bibitem [{\citenamefont {Hsu}\ and\ \citenamefont
  {Sikivie}(1994)}]{HsuPRD1994}%
  \BibitemOpen
  \bibfield  {author} {\bibinfo {author} {\bibfnamefont {S.~D.~H.}\
  \bibnamefont {Hsu}}\ and\ \bibinfo {author} {\bibfnamefont {P.}~\bibnamefont
  {Sikivie}},\ }\bibfield  {title} {\bibinfo {title} {Long-range forces from
  two-neutrino exchange reexamined},\ }\href
  {https://doi.org/10.1103/physrevd.49.4951} {\bibfield  {journal} {\bibinfo
  {journal} {Phys. Rev. D}\ }\textbf {\bibinfo {volume} {49}},\ \bibinfo
  {pages} {4951} (\bibinfo {year} {1994})}\BibitemShut {NoStop}%
\bibitem [{\citenamefont {Kapner}\ \emph {et~al.}(2007)\citenamefont {Kapner},
  \citenamefont {Cook}, \citenamefont {Adelberger}, \citenamefont {Gundlach},
  \citenamefont {Heckel}, \citenamefont {Hoyle},\ and\ \citenamefont
  {Swanson}}]{Kapner2007}%
  \BibitemOpen
  \bibfield  {author} {\bibinfo {author} {\bibfnamefont {D.~J.}\ \bibnamefont
  {Kapner}}, \bibinfo {author} {\bibfnamefont {T.~S.}\ \bibnamefont {Cook}},
  \bibinfo {author} {\bibfnamefont {E.~G.}\ \bibnamefont {Adelberger}},
  \bibinfo {author} {\bibfnamefont {J.~H.}\ \bibnamefont {Gundlach}}, \bibinfo
  {author} {\bibfnamefont {B.~R.}\ \bibnamefont {Heckel}}, \bibinfo {author}
  {\bibfnamefont {C.~D.}\ \bibnamefont {Hoyle}},\ and\ \bibinfo {author}
  {\bibfnamefont {H.~E.}\ \bibnamefont {Swanson}},\ }\bibfield  {title}
  {\bibinfo {title} {{Tests of the gravitational inverse-square law below the
  dark-energy length scale}},\ }\href
  {https://doi.org/10.1103/PhysRevLett.98.021101} {\bibfield  {journal}
  {\bibinfo  {journal} {Phys. Rev. Lett.}\ }\textbf {\bibinfo {volume} {98}},\
  \bibinfo {pages} {021101} (\bibinfo {year} {2007})},\ \Eprint
  {https://arxiv.org/abs/hep-ph/0611184} {arXiv:hep-ph/0611184} \BibitemShut
  {NoStop}%
\bibitem [{\citenamefont {Adelberger}\ \emph {et~al.}(2007)\citenamefont
  {Adelberger}, \citenamefont {Heckel}, \citenamefont {Hoedl}, \citenamefont
  {Hoyle}, \citenamefont {Kapner},\ and\ \citenamefont
  {Upadhye}}]{Adelberger2007}%
  \BibitemOpen
  \bibfield  {author} {\bibinfo {author} {\bibfnamefont {E.~G.}\ \bibnamefont
  {Adelberger}}, \bibinfo {author} {\bibfnamefont {B.~R.}\ \bibnamefont
  {Heckel}}, \bibinfo {author} {\bibfnamefont {S.~A.}\ \bibnamefont {Hoedl}},
  \bibinfo {author} {\bibfnamefont {C.~D.}\ \bibnamefont {Hoyle}}, \bibinfo
  {author} {\bibfnamefont {D.~J.}\ \bibnamefont {Kapner}},\ and\ \bibinfo
  {author} {\bibfnamefont {A.}~\bibnamefont {Upadhye}},\ }\bibfield  {title}
  {\bibinfo {title} {{Particle physics implications of a recent test of the
  gravitational inverse square law}},\ }\href
  {https://doi.org/10.1103/PhysRevLett.98.131104} {\bibfield  {journal}
  {\bibinfo  {journal} {Phys. Rev. Lett.}\ }\textbf {\bibinfo {volume} {98}},\
  \bibinfo {pages} {131104} (\bibinfo {year} {2007})},\ \Eprint
  {https://arxiv.org/abs/hep-ph/0611223} {arXiv:hep-ph/0611223} \BibitemShut
  {NoStop}%
\bibitem [{\citenamefont {Chen}\ \emph {et~al.}(2016)\citenamefont {Chen},
  \citenamefont {Tham}, \citenamefont {Krause}, \citenamefont {Lopez},
  \citenamefont {Fischbach},\ and\ \citenamefont {Decca}}]{Chen2016}%
  \BibitemOpen
  \bibfield  {author} {\bibinfo {author} {\bibfnamefont {Y.~J.}\ \bibnamefont
  {Chen}}, \bibinfo {author} {\bibfnamefont {W.~K.}\ \bibnamefont {Tham}},
  \bibinfo {author} {\bibfnamefont {D.~E.}\ \bibnamefont {Krause}}, \bibinfo
  {author} {\bibfnamefont {D.}~\bibnamefont {Lopez}}, \bibinfo {author}
  {\bibfnamefont {E.}~\bibnamefont {Fischbach}},\ and\ \bibinfo {author}
  {\bibfnamefont {R.~S.}\ \bibnamefont {Decca}},\ }\bibfield  {title} {\bibinfo
  {title} {{Stronger limits on hypothetical Yukawa interactions in the
  30{\textendash}8000 nm range}},\ }\href
  {https://doi.org/10.1103/PhysRevLett.116.221102} {\bibfield  {journal}
  {\bibinfo  {journal} {Phys. Rev. Lett.}\ }\textbf {\bibinfo {volume} {116}},\
  \bibinfo {pages} {221102} (\bibinfo {year} {2016})},\ \Eprint
  {https://arxiv.org/abs/1410.7267} {arXiv:1410.7267 [hep-ex]} \BibitemShut
  {NoStop}%
\bibitem [{\citenamefont {Vasilakis}\ \emph {et~al.}(2009)\citenamefont
  {Vasilakis}, \citenamefont {Brown}, \citenamefont {Kornack},\ and\
  \citenamefont {Romalis}}]{Vasilakis2009}%
  \BibitemOpen
  \bibfield  {author} {\bibinfo {author} {\bibfnamefont {G.}~\bibnamefont
  {Vasilakis}}, \bibinfo {author} {\bibfnamefont {J.~M.}\ \bibnamefont
  {Brown}}, \bibinfo {author} {\bibfnamefont {T.~W.}\ \bibnamefont {Kornack}},\
  and\ \bibinfo {author} {\bibfnamefont {M.~V.}\ \bibnamefont {Romalis}},\
  }\bibfield  {title} {\bibinfo {title} {{Limits on new long range nuclear
  spin-dependent forces set with a K--$^3$He comagnetometer}},\ }\href
  {https://doi.org/10.1103/PhysRevLett.103.261801} {\bibfield  {journal}
  {\bibinfo  {journal} {Phys. Rev. Lett.}\ }\textbf {\bibinfo {volume} {103}},\
  \bibinfo {pages} {261801} (\bibinfo {year} {2009})},\ \Eprint
  {https://arxiv.org/abs/0809.4700} {arXiv:0809.4700 [physics.atom-ph]}
  \BibitemShut {NoStop}%
\bibitem [{\citenamefont {Terrano}\ \emph {et~al.}(2015)\citenamefont
  {Terrano}, \citenamefont {Adelberger}, \citenamefont {Lee},\ and\
  \citenamefont {Heckel}}]{Terrano2015}%
  \BibitemOpen
  \bibfield  {author} {\bibinfo {author} {\bibfnamefont {W.~A.}\ \bibnamefont
  {Terrano}}, \bibinfo {author} {\bibfnamefont {E.~G.}\ \bibnamefont
  {Adelberger}}, \bibinfo {author} {\bibfnamefont {J.~G.}\ \bibnamefont
  {Lee}},\ and\ \bibinfo {author} {\bibfnamefont {B.~R.}\ \bibnamefont
  {Heckel}},\ }\bibfield  {title} {\bibinfo {title} {{Short-range
  spin-dependent interactions of electrons: a probe for exotic pseudo-Goldstone
  bosons}},\ }\href {https://doi.org/10.1103/PhysRevLett.115.201801} {\bibfield
   {journal} {\bibinfo  {journal} {Phys. Rev. Lett.}\ }\textbf {\bibinfo
  {volume} {115}},\ \bibinfo {pages} {201801} (\bibinfo {year} {2015})},\
  \Eprint {https://arxiv.org/abs/1508.02463} {arXiv:1508.02463 [hep-ex]}
  \BibitemShut {NoStop}%
\bibitem [{\citenamefont {Stadnik}(2018)}]{StadnikPRL2018}%
  \BibitemOpen
  \bibfield  {author} {\bibinfo {author} {\bibfnamefont {Y.~V.}\ \bibnamefont
  {Stadnik}},\ }\bibfield  {title} {\bibinfo {title} {Probing long-range
  neutrino-mediated forces with atomic and nuclear spectroscopy},\ }\href
  {https://doi.org/10.1103/PhysRevLett.120.223202} {\bibfield  {journal}
  {\bibinfo  {journal} {Phys. Rev. Lett.}\ }\textbf {\bibinfo {volume} {120}},\
  \bibinfo {pages} {223202} (\bibinfo {year} {2018})}\BibitemShut {NoStop}%
\bibitem [{\citenamefont {Stadnik}\ \emph {et~al.}(2022)\citenamefont
  {Stadnik}, \citenamefont {Dzuba}, \citenamefont {Flambaum},\ and\
  \citenamefont {Munro-Laylim}}]{StadnikErrata}%
  \BibitemOpen
  \bibfield  {author} {\bibinfo {author} {\bibfnamefont {Y.~V.}\ \bibnamefont
  {Stadnik}}, \bibinfo {author} {\bibfnamefont {V.~V.}\ \bibnamefont {Dzuba}},
  \bibinfo {author} {\bibfnamefont {V.~V.}\ \bibnamefont {Flambaum}},\ and\
  \bibinfo {author} {\bibfnamefont {P.}~\bibnamefont {Munro-Laylim}},\
  }\bibfield  {title} {\bibinfo {title} {Errata: Probing long-range
  neutrino-mediated forces with atomic and nuclear spectroscopy},\ }\href
  {https://doi.org/10.1103/PhysRevLett.120.223202} {\bibfield  {journal}
  {\bibinfo  {journal} {Phys. Rev. Lett.}\ }\textbf {\bibinfo {volume} {129}},\
  \bibinfo {pages} {239901} (\bibinfo {year} {2022})}\BibitemShut {NoStop}%
\bibitem [{\citenamefont {Munro-Laylim}\ \emph {et~al.}(2023)\citenamefont
  {Munro-Laylim}, \citenamefont {Dzuba},\ and\ \citenamefont
  {Flambaum}}]{Munro}%
  \BibitemOpen
  \bibfield  {author} {\bibinfo {author} {\bibfnamefont {P.}~\bibnamefont
  {Munro-Laylim}}, \bibinfo {author} {\bibfnamefont {V.}~\bibnamefont
  {Dzuba}},\ and\ \bibinfo {author} {\bibfnamefont {V.}~\bibnamefont
  {Flambaum}},\ }\bibfield  {title} {\bibinfo {title} {Effects of the
  long-range neutrino-mediated force in atomic phenomena},\ }\href
  {https://doi.org/10.1080/00268976.2022.2160385} {\bibfield  {journal}
  {\bibinfo  {journal} {Mol. Phys.}\ }\textbf {\bibinfo {volume} {121}},\
  \bibinfo {pages} {e2160385} (\bibinfo {year} {2023})}\BibitemShut {NoStop}%
\bibitem [{\citenamefont {Flambaum}\ and\ \citenamefont
  {Shuryak}(2007)}]{FS07}%
  \BibitemOpen
  \bibfield  {author} {\bibinfo {author} {\bibfnamefont {V.~V.}\ \bibnamefont
  {Flambaum}}\ and\ \bibinfo {author} {\bibfnamefont {E.~V.}\ \bibnamefont
  {Shuryak}},\ }\bibfield  {title} {\bibinfo {title} {{Influence of meson's
  widths on Yukawa-like potentials and lattice correlation functions}},\ }\href
  {https://doi.org/10.1103/PhysRevC.76.065206} {\bibfield  {journal} {\bibinfo
  {journal} {Phys. Rev. C}\ }\textbf {\bibinfo {volume} {76}},\ \bibinfo
  {pages} {065206} (\bibinfo {year} {2007})},\ \Eprint
  {https://arxiv.org/abs/nucl-th/0702038} {arXiv:nucl-th/0702038} \BibitemShut
  {NoStop}%
\bibitem [{\citenamefont {Ghosh}\ \emph {et~al.}(2020)\citenamefont {Ghosh},
  \citenamefont {Grossman},\ and\ \citenamefont {Tangarife}}]{Ghosh}%
  \BibitemOpen
  \bibfield  {author} {\bibinfo {author} {\bibfnamefont {M.}~\bibnamefont
  {Ghosh}}, \bibinfo {author} {\bibfnamefont {Y.}~\bibnamefont {Grossman}},\
  and\ \bibinfo {author} {\bibfnamefont {W.}~\bibnamefont {Tangarife}},\
  }\bibfield  {title} {\bibinfo {title} {{Probing the two-neutrino exchange
  force using atomic parity violation}},\ }\href
  {https://doi.org/10.1103/PhysRevD.101.116006} {\bibfield  {journal} {\bibinfo
   {journal} {Phys. Rev. D}\ }\textbf {\bibinfo {volume} {101}},\ \bibinfo
  {pages} {116006} (\bibinfo {year} {2020})},\ \Eprint
  {https://arxiv.org/abs/1912.09444} {arXiv:1912.09444 [hep-ph]} \BibitemShut
  {NoStop}%
\bibitem [{\citenamefont {Dzuba}\ \emph {et~al.}(2022)\citenamefont {Dzuba},
  \citenamefont {Flambaum},\ and\ \citenamefont {Munro-Laylim}}]{MunroPV}%
  \BibitemOpen
  \bibfield  {author} {\bibinfo {author} {\bibfnamefont {V.~A.}\ \bibnamefont
  {Dzuba}}, \bibinfo {author} {\bibfnamefont {V.~V.}\ \bibnamefont
  {Flambaum}},\ and\ \bibinfo {author} {\bibfnamefont {P.}~\bibnamefont
  {Munro-Laylim}},\ }\bibfield  {title} {\bibinfo {title} {{Long-range
  parity-nonconserving electron-nucleon interaction}},\ }\href
  {https://doi.org/10.1103/PhysRevA.106.012817} {\bibfield  {journal} {\bibinfo
   {journal} {Phys. Rev. A}\ }\textbf {\bibinfo {volume} {106}},\ \bibinfo
  {pages} {012817} (\bibinfo {year} {2022})},\ \Eprint
  {https://arxiv.org/abs/2205.02569} {arXiv:2205.02569 [hep-ph]} \BibitemShut
  {NoStop}%
\bibitem [{\citenamefont {Ghosh}\ \emph
  {et~al.}(2025{\natexlab{a}})\citenamefont {Ghosh}, \citenamefont {Grossman},
  \citenamefont {Sieng},\ and\ \citenamefont {Yu}}]{Ghosh:2024ctv}%
  \BibitemOpen
  \bibfield  {author} {\bibinfo {author} {\bibfnamefont {M.}~\bibnamefont
  {Ghosh}}, \bibinfo {author} {\bibfnamefont {Y.}~\bibnamefont {Grossman}},
  \bibinfo {author} {\bibfnamefont {C.}~\bibnamefont {Sieng}},\ and\ \bibinfo
  {author} {\bibfnamefont {B.}~\bibnamefont {Yu}},\ }\bibfield  {title}
  {\bibinfo {title} {{Neutrino force at all length scales}},\ }\href
  {https://doi.org/10.1103/PhysRevD.112.113004} {\bibfield  {journal} {\bibinfo
   {journal} {Phys. Rev. D}\ }\textbf {\bibinfo {volume} {112}},\ \bibinfo
  {pages} {113004} (\bibinfo {year} {2025}{\natexlab{a}})},\ \Eprint
  {https://arxiv.org/abs/2410.19059} {arXiv:2410.19059 [hep-ph]} \BibitemShut
  {NoStop}%
\bibitem [{\citenamefont {Ghosh}\ \emph
  {et~al.}(2025{\natexlab{b}})\citenamefont {Ghosh}, \citenamefont {Grossman},
  \citenamefont {Sieng},\ and\ \citenamefont {Yu}}]{Ghosh:2025ole}%
  \BibitemOpen
  \bibfield  {author} {\bibinfo {author} {\bibfnamefont {M.}~\bibnamefont
  {Ghosh}}, \bibinfo {author} {\bibfnamefont {Y.}~\bibnamefont {Grossman}},
  \bibinfo {author} {\bibfnamefont {C.}~\bibnamefont {Sieng}},\ and\ \bibinfo
  {author} {\bibfnamefont {B.}~\bibnamefont {Yu}},\ }\bibfield  {title}
  {\bibinfo {title} {{Neutrino effects on atomic measurements of the Weinberg
  angle}},\ }\href {https://doi.org/10.1103/g7tx-srnn} {\bibfield  {journal}
  {\bibinfo  {journal} {Phys. Rev. Lett.}\ }\textbf {\bibinfo {volume} {135}},\
  \bibinfo {pages} {251803} (\bibinfo {year} {2025}{\natexlab{b}})},\ \Eprint
  {https://arxiv.org/abs/2512.07938} {arXiv:2512.07938 [hep-ph]} \BibitemShut
  {NoStop}%
\bibitem [{\citenamefont {Wood}\ \emph {et~al.}(1997)\citenamefont {Wood},
  \citenamefont {Bennett}, \citenamefont {Cho}, \citenamefont {Masterson},
  \citenamefont {Roberts}, \citenamefont {Tanner},\ and\ \citenamefont
  {Wieman}}]{Wood}%
  \BibitemOpen
  \bibfield  {author} {\bibinfo {author} {\bibfnamefont {C.~S.}\ \bibnamefont
  {Wood}}, \bibinfo {author} {\bibfnamefont {S.~C.}\ \bibnamefont {Bennett}},
  \bibinfo {author} {\bibfnamefont {D.}~\bibnamefont {Cho}}, \bibinfo {author}
  {\bibfnamefont {B.~P.}\ \bibnamefont {Masterson}}, \bibinfo {author}
  {\bibfnamefont {J.~L.}\ \bibnamefont {Roberts}}, \bibinfo {author}
  {\bibfnamefont {C.~E.}\ \bibnamefont {Tanner}},\ and\ \bibinfo {author}
  {\bibfnamefont {C.~E.}\ \bibnamefont {Wieman}},\ }\bibfield  {title}
  {\bibinfo {title} {{Measurement of parity nonconservation and an anapole
  moment in cesium}},\ }\href {https://doi.org/10.1126/science.275.5307.1759}
  {\bibfield  {journal} {\bibinfo  {journal} {Science}\ }\textbf {\bibinfo
  {volume} {275}},\ \bibinfo {pages} {1759} (\bibinfo {year}
  {1997})}\BibitemShut {NoStop}%
\bibitem [{\citenamefont {Dzuba}\ \emph {et~al.}(1989)\citenamefont {Dzuba},
  \citenamefont {Flambaum},\ and\ \citenamefont {Sushkov}}]{DzuFlaSus89a}%
  \BibitemOpen
  \bibfield  {author} {\bibinfo {author} {\bibfnamefont {V.~A.}\ \bibnamefont
  {Dzuba}}, \bibinfo {author} {\bibfnamefont {V.~V.}\ \bibnamefont
  {Flambaum}},\ and\ \bibinfo {author} {\bibfnamefont {O.~P.}\ \bibnamefont
  {Sushkov}},\ }\bibfield  {title} {\bibinfo {title} {{Summation of the high
  orders of perturbation theory for the parity nonconserving E1 amplitude of
  6s-7s transition in cesium atom}},\ }\href
  {https://doi.org/10.1016/0375-9601(89)90777-9} {\bibfield  {journal}
  {\bibinfo  {journal} {Phys. Lett. A}\ }\textbf {\bibinfo {volume} {141}},\
  \bibinfo {pages} {147} (\bibinfo {year} {1989})}\BibitemShut {NoStop}%
\bibitem [{\citenamefont {Dzuba}\ \emph {et~al.}(2002)\citenamefont {Dzuba},
  \citenamefont {Flambaum},\ and\ \citenamefont {Ginges}}]{DzuFlaGin02}%
  \BibitemOpen
  \bibfield  {author} {\bibinfo {author} {\bibfnamefont {V.~A.}\ \bibnamefont
  {Dzuba}}, \bibinfo {author} {\bibfnamefont {V.~V.}\ \bibnamefont
  {Flambaum}},\ and\ \bibinfo {author} {\bibfnamefont {J.~S.~M.}\ \bibnamefont
  {Ginges}},\ }\bibfield  {title} {\bibinfo {title} {{High-precision
  calculation of parity nonconservation in cesium and test of the standard
  model}},\ }\href {https://doi.org/10.1103/PhysRevD.66.076013} {\bibfield
  {journal} {\bibinfo  {journal} {Phys. Rev. D}\ }\textbf {\bibinfo {volume}
  {66}},\ \bibinfo {pages} {076013} (\bibinfo {year} {2002})},\ \Eprint
  {https://arxiv.org/abs/hep-ph/0204134} {arXiv:hep-ph/0204134} \BibitemShut
  {NoStop}%
\bibitem [{\citenamefont {Blundell}\ \emph {et~al.}(1990)\citenamefont
  {Blundell}, \citenamefont {Johnson},\ and\ \citenamefont
  {Sapirstein}}]{Blundell90}%
  \BibitemOpen
  \bibfield  {author} {\bibinfo {author} {\bibfnamefont {S.~A.}\ \bibnamefont
  {Blundell}}, \bibinfo {author} {\bibfnamefont {W.~R.}\ \bibnamefont
  {Johnson}},\ and\ \bibinfo {author} {\bibfnamefont {J.}~\bibnamefont
  {Sapirstein}},\ }\bibfield  {title} {\bibinfo {title} {{High accuracy
  calculation of the $6s_{1/2}\to7s_{1/2}$ parity nonconserving transition in
  atomic cesium and implications for the standard model}},\ }\href
  {https://doi.org/10.1103/PhysRevLett.65.1411} {\bibfield  {journal} {\bibinfo
   {journal} {Phys. Rev. Lett.}\ }\textbf {\bibinfo {volume} {65}},\ \bibinfo
  {pages} {1411} (\bibinfo {year} {1990})}\BibitemShut {NoStop}%
\bibitem [{\citenamefont {Blundell}\ \emph {et~al.}(1992)\citenamefont
  {Blundell}, \citenamefont {Sapirstein},\ and\ \citenamefont
  {Johnson}}]{Blundell92}%
  \BibitemOpen
  \bibfield  {author} {\bibinfo {author} {\bibfnamefont {S.~A.}\ \bibnamefont
  {Blundell}}, \bibinfo {author} {\bibfnamefont {J.~R.}\ \bibnamefont
  {Sapirstein}},\ and\ \bibinfo {author} {\bibfnamefont {W.~R.}\ \bibnamefont
  {Johnson}},\ }\bibfield  {title} {\bibinfo {title} {{High accuracy
  calculation of parity nonconservation in cesium and implications for particle
  physics}},\ }\href {https://doi.org/10.1103/PhysRevD.45.1602} {\bibfield
  {journal} {\bibinfo  {journal} {Phys. Rev. D}\ }\textbf {\bibinfo {volume}
  {45}},\ \bibinfo {pages} {1602} (\bibinfo {year} {1992})}\BibitemShut
  {NoStop}%
\bibitem [{\citenamefont {Flambaum}\ and\ \citenamefont
  {Ginges}(2005)}]{Ginges05}%
  \BibitemOpen
  \bibfield  {author} {\bibinfo {author} {\bibfnamefont {V.~V.}\ \bibnamefont
  {Flambaum}}\ and\ \bibinfo {author} {\bibfnamefont {J.~S.~M.}\ \bibnamefont
  {Ginges}},\ }\bibfield  {title} {\bibinfo {title} {{The radiative potential
  method for calculations of QED radiative corrections to energy levels and
  electromagnetic amplitudes in many-electron atoms}},\ }\href
  {https://doi.org/10.1103/PhysRevA.72.052115} {\bibfield  {journal} {\bibinfo
  {journal} {Phys. Rev. A}\ }\textbf {\bibinfo {volume} {72}},\ \bibinfo
  {pages} {052115} (\bibinfo {year} {2005})},\ \Eprint
  {https://arxiv.org/abs/physics/0507067} {arXiv:physics/0507067} \BibitemShut
  {NoStop}%
\bibitem [{\citenamefont {Porsev}\ \emph {et~al.}(2009)\citenamefont {Porsev},
  \citenamefont {Beloy},\ and\ \citenamefont {Derevianko}}]{PBD09}%
  \BibitemOpen
  \bibfield  {author} {\bibinfo {author} {\bibfnamefont {S.~G.}\ \bibnamefont
  {Porsev}}, \bibinfo {author} {\bibfnamefont {K.}~\bibnamefont {Beloy}},\ and\
  \bibinfo {author} {\bibfnamefont {A.}~\bibnamefont {Derevianko}},\ }\bibfield
   {title} {\bibinfo {title} {{Precision determination of electroweak coupling
  from atomic parity violation and implications for particle physics}},\ }\href
  {https://doi.org/10.1103/PhysRevLett.102.181601} {\bibfield  {journal}
  {\bibinfo  {journal} {Phys. Rev. Lett.}\ }\textbf {\bibinfo {volume} {102}},\
  \bibinfo {pages} {181601} (\bibinfo {year} {2009})},\ \Eprint
  {https://arxiv.org/abs/0902.0335} {arXiv:0902.0335 [hep-ph]} \BibitemShut
  {NoStop}%
\bibitem [{\citenamefont {Porsev}\ \emph {et~al.}(2010)\citenamefont {Porsev},
  \citenamefont {Beloy},\ and\ \citenamefont {Derevianko}}]{PBD10}%
  \BibitemOpen
  \bibfield  {author} {\bibinfo {author} {\bibfnamefont {S.~G.}\ \bibnamefont
  {Porsev}}, \bibinfo {author} {\bibfnamefont {K.}~\bibnamefont {Beloy}},\ and\
  \bibinfo {author} {\bibfnamefont {A.}~\bibnamefont {Derevianko}},\ }\bibfield
   {title} {\bibinfo {title} {{Precision determination of weak charge of
  $^{133}$Cs from atomic parity violation}},\ }\href
  {https://doi.org/10.1103/PhysRevD.82.036008} {\bibfield  {journal} {\bibinfo
  {journal} {Phys. Rev. D}\ }\textbf {\bibinfo {volume} {82}},\ \bibinfo
  {pages} {036008} (\bibinfo {year} {2010})},\ \Eprint
  {https://arxiv.org/abs/1006.4193} {arXiv:1006.4193 [hep-ph]} \BibitemShut
  {NoStop}%
\bibitem [{\citenamefont {Dzuba}\ \emph {et~al.}(2012)\citenamefont {Dzuba},
  \citenamefont {Berengut}, \citenamefont {Flambaum},\ and\ \citenamefont
  {Roberts}}]{CsPNC12}%
  \BibitemOpen
  \bibfield  {author} {\bibinfo {author} {\bibfnamefont {V.~A.}\ \bibnamefont
  {Dzuba}}, \bibinfo {author} {\bibfnamefont {J.~C.}\ \bibnamefont {Berengut}},
  \bibinfo {author} {\bibfnamefont {V.~V.}\ \bibnamefont {Flambaum}},\ and\
  \bibinfo {author} {\bibfnamefont {B.}~\bibnamefont {Roberts}},\ }\bibfield
  {title} {\bibinfo {title} {{Revisiting parity non-conservation in cesium}},\
  }\href {https://doi.org/10.1103/PhysRevLett.109.203003} {\bibfield  {journal}
  {\bibinfo  {journal} {Phys. Rev. Lett.}\ }\textbf {\bibinfo {volume} {109}},\
  \bibinfo {pages} {203003} (\bibinfo {year} {2012})},\ \Eprint
  {https://arxiv.org/abs/1207.5864} {arXiv:1207.5864 [hep-ph]} \BibitemShut
  {NoStop}%
\bibitem [{\citenamefont {Damitz}\ \emph {et~al.}(2019)\citenamefont {Damitz},
  \citenamefont {Toh}, \citenamefont {Putney}, \citenamefont {Tanner},\ and\
  \citenamefont {Elliott}}]{Elliott1}%
  \BibitemOpen
  \bibfield  {author} {\bibinfo {author} {\bibfnamefont {A.}~\bibnamefont
  {Damitz}}, \bibinfo {author} {\bibfnamefont {G.}~\bibnamefont {Toh}},
  \bibinfo {author} {\bibfnamefont {E.}~\bibnamefont {Putney}}, \bibinfo
  {author} {\bibfnamefont {C.~E.}\ \bibnamefont {Tanner}},\ and\ \bibinfo
  {author} {\bibfnamefont {D.~S.}\ \bibnamefont {Elliott}},\ }\bibfield
  {title} {\bibinfo {title} {{Measurement of the radial matrix elements for the
  $6s~^2S_{1/2}\to 7p~^2P_J$ transitions in cesium}},\ }\href
  {https://doi.org/10.1103/PhysRevA.99.062510} {\bibfield  {journal} {\bibinfo
  {journal} {Phys. Rev. A}\ }\textbf {\bibinfo {volume} {99}},\ \bibinfo
  {pages} {062510} (\bibinfo {year} {2019})},\ \Eprint
  {https://arxiv.org/abs/1904.06362} {arXiv:1904.06362 [physics.atom-ph]}
  \BibitemShut {NoStop}%
\bibitem [{\citenamefont {Toh}\ \emph {et~al.}(2019)\citenamefont {Toh},
  \citenamefont {Damitz}, \citenamefont {Glotzbach}, \citenamefont {Quirk},
  \citenamefont {Stevenson}, \citenamefont {Choi}, \citenamefont {Safronova},\
  and\ \citenamefont {Elliott}}]{Elliott2}%
  \BibitemOpen
  \bibfield  {author} {\bibinfo {author} {\bibfnamefont {G.}~\bibnamefont
  {Toh}}, \bibinfo {author} {\bibfnamefont {A.}~\bibnamefont {Damitz}},
  \bibinfo {author} {\bibfnamefont {N.}~\bibnamefont {Glotzbach}}, \bibinfo
  {author} {\bibfnamefont {J.}~\bibnamefont {Quirk}}, \bibinfo {author}
  {\bibfnamefont {I.~C.}\ \bibnamefont {Stevenson}}, \bibinfo {author}
  {\bibfnamefont {J.}~\bibnamefont {Choi}}, \bibinfo {author} {\bibfnamefont
  {M.~S.}\ \bibnamefont {Safronova}},\ and\ \bibinfo {author} {\bibfnamefont
  {D.~S.}\ \bibnamefont {Elliott}},\ }\bibfield  {title} {\bibinfo {title}
  {{Electric dipole matrix elements for the
  $6p{\phantom{\rule{0.16em}{0ex}}}^{2}{P}_{J}\ensuremath{\rightarrow}7s{\phantom{\rule{0.16em}{0ex}}}^{2}{S}_{1/2}$
  transition in atomic cesium}},\ }\href
  {https://doi.org/10.1103/PhysRevA.99.032504} {\bibfield  {journal} {\bibinfo
  {journal} {Phys. Rev. A}\ }\textbf {\bibinfo {volume} {99}},\ \bibinfo
  {pages} {032504} (\bibinfo {year} {2019})},\ \Eprint
  {https://arxiv.org/abs/1901.00480} {arXiv:1901.00480 [physics.atom-ph]}
  \BibitemShut {NoStop}%
\bibitem [{\citenamefont {Quirk}\ \emph {et~al.}(2022)\citenamefont {Quirk},
  \citenamefont {Sherman}, \citenamefont {Damitz}, \citenamefont {Tanner},\
  and\ \citenamefont {Elliott}}]{Elliott3}%
  \BibitemOpen
  \bibfield  {author} {\bibinfo {author} {\bibfnamefont {J.~A.}\ \bibnamefont
  {Quirk}}, \bibinfo {author} {\bibfnamefont {L.}~\bibnamefont {Sherman}},
  \bibinfo {author} {\bibfnamefont {A.}~\bibnamefont {Damitz}}, \bibinfo
  {author} {\bibfnamefont {C.~E.}\ \bibnamefont {Tanner}},\ and\ \bibinfo
  {author} {\bibfnamefont {D.~S.}\ \bibnamefont {Elliott}},\ }\bibfield
  {title} {\bibinfo {title} {{Measurement of the hyperfine coupling constants
  and absolute energies of the $12s {}^{2}{S}_{1/2}, 13s {}^{2}{S}_{1/2}$, and
  $11d {}^{2}{D}_{J}$ levels in atomic cesium}},\ }\href
  {https://doi.org/10.1103/PhysRevA.105.022819} {\bibfield  {journal} {\bibinfo
   {journal} {Phys. Rev. A}\ }\textbf {\bibinfo {volume} {105}},\ \bibinfo
  {pages} {022819} (\bibinfo {year} {2022})},\ \Eprint
  {https://arxiv.org/abs/2201.00755} {arXiv:2201.00755 [physics.atom-ph]}
  \BibitemShut {NoStop}%
\bibitem [{\citenamefont {Tan}\ \emph {et~al.}(2022)\citenamefont {Tan},
  \citenamefont {Xiao},\ and\ \citenamefont {Derevianko}}]{TXD22}%
  \BibitemOpen
  \bibfield  {author} {\bibinfo {author} {\bibfnamefont {H.~B.~T.}\
  \bibnamefont {Tan}}, \bibinfo {author} {\bibfnamefont {D.}~\bibnamefont
  {Xiao}},\ and\ \bibinfo {author} {\bibfnamefont {A.}~\bibnamefont
  {Derevianko}},\ }\bibfield  {title} {\bibinfo {title} {{Parity-mixed
  coupled-cluster formalism for computing parity-violating amplitudes}},\
  }\href {https://doi.org/10.1103/PhysRevA.105.022803} {\bibfield  {journal}
  {\bibinfo  {journal} {Phys. Rev. A}\ }\textbf {\bibinfo {volume} {105}},\
  \bibinfo {pages} {022803} (\bibinfo {year} {2022})},\ \Eprint
  {https://arxiv.org/abs/2112.04059} {arXiv:2112.04059 [physics.atom-ph]}
  \BibitemShut {NoStop}%
\bibitem [{\citenamefont {Dzuba}\ \emph {et~al.}(1986)\citenamefont {Dzuba},
  \citenamefont {Flambaum},\ and\ \citenamefont {Khriplovich}}]{DzuFlaKhr86}%
  \BibitemOpen
  \bibfield  {author} {\bibinfo {author} {\bibfnamefont {V.~A.}\ \bibnamefont
  {Dzuba}}, \bibinfo {author} {\bibfnamefont {V.~V.}\ \bibnamefont
  {Flambaum}},\ and\ \bibinfo {author} {\bibfnamefont {I.~B.}\ \bibnamefont
  {Khriplovich}},\ }\bibfield  {title} {\bibinfo {title} {{Enhancement of $P$-
  and $T$-nonconserving effects in rare-earth atoms}},\ }\href
  {https://doi.org/10.1007/BF01436678} {\bibfield  {journal} {\bibinfo
  {journal} {Z. Phys. D}\ }\textbf {\bibinfo {volume} {1}},\ \bibinfo {pages}
  {243} (\bibinfo {year} {1986})}\BibitemShut {NoStop}%
\bibitem [{\citenamefont {Brown}\ \emph {et~al.}(2009)\citenamefont {Brown},
  \citenamefont {Derevianko},\ and\ \citenamefont {Flambaum}}]{BDF09}%
  \BibitemOpen
  \bibfield  {author} {\bibinfo {author} {\bibfnamefont {B.~A.}\ \bibnamefont
  {Brown}}, \bibinfo {author} {\bibfnamefont {A.}~\bibnamefont {Derevianko}},\
  and\ \bibinfo {author} {\bibfnamefont {V.~V.}\ \bibnamefont {Flambaum}},\
  }\bibfield  {title} {\bibinfo {title} {{Calculations of the neutron skin and
  its effect in atomic parity violation}},\ }\href
  {https://doi.org/10.1103/PhysRevC.79.035501} {\bibfield  {journal} {\bibinfo
  {journal} {Phys. Rev. C}\ }\textbf {\bibinfo {volume} {79}},\ \bibinfo
  {pages} {035501} (\bibinfo {year} {2009})},\ \Eprint
  {https://arxiv.org/abs/0804.4315} {arXiv:0804.4315 [hep-ph]} \BibitemShut
  {NoStop}%
\bibitem [{\citenamefont {Viatkina}\ \emph {et~al.}(2019)\citenamefont
  {Viatkina}, \citenamefont {Antypas}, \citenamefont {Kozlov}, \citenamefont
  {Budker},\ and\ \citenamefont {Flambaum}}]{VF19}%
  \BibitemOpen
  \bibfield  {author} {\bibinfo {author} {\bibfnamefont {A.~V.}\ \bibnamefont
  {Viatkina}}, \bibinfo {author} {\bibfnamefont {D.}~\bibnamefont {Antypas}},
  \bibinfo {author} {\bibfnamefont {M.~G.}\ \bibnamefont {Kozlov}}, \bibinfo
  {author} {\bibfnamefont {D.}~\bibnamefont {Budker}},\ and\ \bibinfo {author}
  {\bibfnamefont {V.~V.}\ \bibnamefont {Flambaum}},\ }\bibfield  {title}
  {\bibinfo {title} {{Dependence of atomic parity-violation effects on neutron
  skins and new physics}},\ }\href
  {https://doi.org/10.1103/PhysRevC.100.034318} {\bibfield  {journal} {\bibinfo
   {journal} {Phys. Rev. C}\ }\textbf {\bibinfo {volume} {100}},\ \bibinfo
  {pages} {034318} (\bibinfo {year} {2019})},\ \Eprint
  {https://arxiv.org/abs/1903.00123} {arXiv:1903.00123 [physics.atom-ph]}
  \BibitemShut {NoStop}%
\bibitem [{\citenamefont {Antypas}\ \emph {et~al.}(2019)\citenamefont
  {Antypas}, \citenamefont {Fabricant}, \citenamefont {Stalnaker},
  \citenamefont {Tsigutkin}, \citenamefont {Flambaum},\ and\ \citenamefont
  {Budker}}]{Budker19}%
  \BibitemOpen
  \bibfield  {author} {\bibinfo {author} {\bibfnamefont {D.}~\bibnamefont
  {Antypas}}, \bibinfo {author} {\bibfnamefont {A.}~\bibnamefont {Fabricant}},
  \bibinfo {author} {\bibfnamefont {J.~E.}\ \bibnamefont {Stalnaker}}, \bibinfo
  {author} {\bibfnamefont {K.}~\bibnamefont {Tsigutkin}}, \bibinfo {author}
  {\bibfnamefont {V.~V.}\ \bibnamefont {Flambaum}},\ and\ \bibinfo {author}
  {\bibfnamefont {D.}~\bibnamefont {Budker}},\ }\bibfield  {title} {\bibinfo
  {title} {{Isotopic variation of parity violation in atomic ytterbium}},\
  }\href {https://doi.org/10.1038/s41567-018-0312-8} {\bibfield  {journal}
  {\bibinfo  {journal} {Nature Phys.}\ }\textbf {\bibinfo {volume} {15}},\
  \bibinfo {pages} {120} (\bibinfo {year} {2019})},\ \Eprint
  {https://arxiv.org/abs/1804.05747} {arXiv:1804.05747 [physics.atom-ph]}
  \BibitemShut {NoStop}%
\bibitem [{\citenamefont {Roberts}\ \emph {et~al.}(2015)\citenamefont
  {Roberts}, \citenamefont {Dzuba},\ and\ \citenamefont {Flambaum}}]{RevPNC}%
  \BibitemOpen
  \bibfield  {author} {\bibinfo {author} {\bibfnamefont {B.~M.}\ \bibnamefont
  {Roberts}}, \bibinfo {author} {\bibfnamefont {V.~A.}\ \bibnamefont {Dzuba}},\
  and\ \bibinfo {author} {\bibfnamefont {V.~V.}\ \bibnamefont {Flambaum}},\
  }\bibfield  {title} {\bibinfo {title} {{Parity and time-reversal violation in
  atomic systems}},\ }\href
  {https://doi.org/10.1146/annurev-nucl-102014-022331} {\bibfield  {journal}
  {\bibinfo  {journal} {Ann. Rev. Nucl. Part. Sci.}\ }\textbf {\bibinfo
  {volume} {65}},\ \bibinfo {pages} {63} (\bibinfo {year} {2015})},\ \Eprint
  {https://arxiv.org/abs/1412.6644} {arXiv:1412.6644 [physics.atom-ph]}
  \BibitemShut {NoStop}%
\bibitem [{\citenamefont {Safronova}\ \emph {et~al.}(2018)\citenamefont
  {Safronova}, \citenamefont {Budker}, \citenamefont {DeMille}, \citenamefont
  {Kimball}, \citenamefont {Derevianko},\ and\ \citenamefont
  {Clark}}]{RevPNC1}%
  \BibitemOpen
  \bibfield  {author} {\bibinfo {author} {\bibfnamefont {M.~S.}\ \bibnamefont
  {Safronova}}, \bibinfo {author} {\bibfnamefont {D.}~\bibnamefont {Budker}},
  \bibinfo {author} {\bibfnamefont {D.}~\bibnamefont {DeMille}}, \bibinfo
  {author} {\bibfnamefont {D.~F.~J.}\ \bibnamefont {Kimball}}, \bibinfo
  {author} {\bibfnamefont {A.}~\bibnamefont {Derevianko}},\ and\ \bibinfo
  {author} {\bibfnamefont {C.~W.}\ \bibnamefont {Clark}},\ }\bibfield  {title}
  {\bibinfo {title} {Search for new physics with atoms and molecules},\ }\href
  {https://doi.org/10.1103/RevModPhys.90.025008} {\bibfield  {journal}
  {\bibinfo  {journal} {Rev. Mod. Phys.}\ }\textbf {\bibinfo {volume} {90}},\
  \bibinfo {pages} {025008} (\bibinfo {year} {2018})},\ \Eprint
  {https://arxiv.org/abs/1710.01833} {arXiv:1710.01833 [physics.atom-ph]}
  \BibitemShut {NoStop}%
\bibitem [{\citenamefont {Marciano}\ and\ \citenamefont
  {Sirlin}(1983)}]{Marciano}%
  \BibitemOpen
  \bibfield  {author} {\bibinfo {author} {\bibfnamefont {W.~J.}\ \bibnamefont
  {Marciano}}\ and\ \bibinfo {author} {\bibfnamefont {A.}~\bibnamefont
  {Sirlin}},\ }\bibfield  {title} {\bibinfo {title} {{Radiative corrections to
  atomic parity violation}},\ }\href {https://doi.org/10.1103/PhysRevD.27.552}
  {\bibfield  {journal} {\bibinfo  {journal} {Phys. Rev. D}\ }\textbf {\bibinfo
  {volume} {27}},\ \bibinfo {pages} {552} (\bibinfo {year} {1983})}\BibitemShut
  {NoStop}%
\bibitem [{\citenamefont {Marciano}\ and\ \citenamefont
  {Sanda}(1978)}]{MarcianoSanda}%
  \BibitemOpen
  \bibfield  {author} {\bibinfo {author} {\bibfnamefont {W.~J.}\ \bibnamefont
  {Marciano}}\ and\ \bibinfo {author} {\bibfnamefont {A.~I.}\ \bibnamefont
  {Sanda}},\ }\bibfield  {title} {\bibinfo {title} {Parity violation in atoms
  induced by radiative corrections},\ }\href
  {https://doi.org/10.1103/PhysRevD.17.3055} {\bibfield  {journal} {\bibinfo
  {journal} {Phys. Rev. D}\ }\textbf {\bibinfo {volume} {17}},\ \bibinfo
  {pages} {3055} (\bibinfo {year} {1978})}\BibitemShut {NoStop}%
\bibitem [{\citenamefont {Marciano}\ and\ \citenamefont
  {Sirlin}(1984)}]{MarcianoSirlin1984}%
  \BibitemOpen
  \bibfield  {author} {\bibinfo {author} {\bibfnamefont {W.~J.}\ \bibnamefont
  {Marciano}}\ and\ \bibinfo {author} {\bibfnamefont {A.}~\bibnamefont
  {Sirlin}},\ }\bibfield  {title} {\bibinfo {title} {{Some general properties
  of the $O(\ensuremath{\alpha})$ corrections to parity violation in atoms}},\
  }\href {https://doi.org/10.1103/PhysRevD.29.75} {\bibfield  {journal}
  {\bibinfo  {journal} {Phys. Rev. D}\ }\textbf {\bibinfo {volume} {29}},\
  \bibinfo {pages} {75} (\bibinfo {year} {1984})}\BibitemShut {NoStop}%
\bibitem [{\citenamefont {Blunden}\ \emph {et~al.}(2012)\citenamefont
  {Blunden}, \citenamefont {Melnitchouk},\ and\ \citenamefont
  {Thomas}}]{Thomas}%
  \BibitemOpen
  \bibfield  {author} {\bibinfo {author} {\bibfnamefont {P.~G.}\ \bibnamefont
  {Blunden}}, \bibinfo {author} {\bibfnamefont {W.}~\bibnamefont
  {Melnitchouk}},\ and\ \bibinfo {author} {\bibfnamefont {A.~W.}\ \bibnamefont
  {Thomas}},\ }\bibfield  {title} {\bibinfo {title} {{$\gamma Z$ box
  corrections to weak charges of heavy nuclei in atomic parity violation}},\
  }\href {https://doi.org/10.1103/PhysRevLett.109.262301} {\bibfield  {journal}
  {\bibinfo  {journal} {Phys. Rev. Lett.}\ }\textbf {\bibinfo {volume} {109}},\
  \bibinfo {pages} {262301} (\bibinfo {year} {2012})}\BibitemShut {NoStop}%
\bibitem [{\citenamefont {Duch}\ \emph {et~al.}(2026)\citenamefont {Duch},
  \citenamefont {Masjuan},\ and\ \citenamefont {Spiesberger}}]{Duch:2026ujd}%
  \BibitemOpen
  \bibfield  {author} {\bibinfo {author} {\bibfnamefont {B.}~\bibnamefont
  {Duch}}, \bibinfo {author} {\bibfnamefont {P.}~\bibnamefont {Masjuan}},\ and\
  \bibinfo {author} {\bibfnamefont {H.}~\bibnamefont {Spiesberger}},\
  }\href@noop {} {\bibinfo {title} {{$\gamma Z$ box at low energy}}} (\bibinfo
  {year} {2026}),\ \Eprint {https://arxiv.org/abs/2601.13819} {arXiv:2601.13819
  [hep-ph]} \BibitemShut {NoStop}%
\bibitem [{\citenamefont {Navas}\ and\ \citenamefont {et.
  al.}(2024)}]{PDG2024RPP}%
  \BibitemOpen
  \bibfield  {author} {\bibinfo {author} {\bibfnamefont {S.}~\bibnamefont
  {Navas}}\ and\ \bibinfo {author} {\bibnamefont {et. al.}},\ }\bibfield
  {title} {\bibinfo {title} {Partcicle data group, review of particle
  physics},\ }\href {https://doi.org/10.1103/PhysRevD.110.030001} {\bibfield
  {journal} {\bibinfo  {journal} {Phys. Rev. D}\ }\textbf {\bibinfo {volume}
  {110}},\ \bibinfo {pages} {030001} (\bibinfo {year} {2024})}\BibitemShut
  {NoStop}%
\bibitem [{\citenamefont {Erler}\ and\ \citenamefont
  {Freitas}(2024)}]{PDG2024SM}%
  \BibitemOpen
  \bibfield  {author} {\bibinfo {author} {\bibfnamefont {J.}~\bibnamefont
  {Erler}}\ and\ \bibinfo {author} {\bibfnamefont {A.}~\bibnamefont
  {Freitas}},\ }\href
  {https://pdg.lbl.gov/2024/reviews/rpp2024-rev-standard-model.pdf} {\bibinfo
  {title} {Electroweak model and constraints on new physics (pdg 2024 review,
  section on atomic parity violation)}},\ \bibinfo {howpublished} {Particle
  Data Group online review PDF} (\bibinfo {year} {2024})\BibitemShut {NoStop}%
\bibitem [{\citenamefont {Tran~Tan}\ \emph {et~al.}(2023)\citenamefont
  {Tran~Tan}, \citenamefont {Xiao},\ and\ \citenamefont
  {Derevianko}}]{BaoBeta}%
  \BibitemOpen
  \bibfield  {author} {\bibinfo {author} {\bibfnamefont {H.~B.}\ \bibnamefont
  {Tran~Tan}}, \bibinfo {author} {\bibfnamefont {D.}~\bibnamefont {Xiao}},\
  and\ \bibinfo {author} {\bibfnamefont {A.}~\bibnamefont {Derevianko}},\
  }\bibfield  {title} {\bibinfo {title} {Reevaluation of stark-induced
  transition polarizabilities in cesium},\ }\href
  {https://doi.org/10.1103/PhysRevA.108.022808} {\bibfield  {journal} {\bibinfo
   {journal} {Phys. Rev. A}\ }\textbf {\bibinfo {volume} {108}},\ \bibinfo
  {pages} {022808} (\bibinfo {year} {2023})}\BibitemShut {NoStop}%
\bibitem [{\citenamefont {Vasilyev}\ \emph {et~al.}(2002)\citenamefont
  {Vasilyev}, \citenamefont {Savukov}, \citenamefont {Safronova},\ and\
  \citenamefont {Berry}}]{Vasil}%
  \BibitemOpen
  \bibfield  {author} {\bibinfo {author} {\bibfnamefont {A.~A.}\ \bibnamefont
  {Vasilyev}}, \bibinfo {author} {\bibfnamefont {I.~M.}\ \bibnamefont
  {Savukov}}, \bibinfo {author} {\bibfnamefont {M.~S.}\ \bibnamefont
  {Safronova}},\ and\ \bibinfo {author} {\bibfnamefont {H.~G.}\ \bibnamefont
  {Berry}},\ }\bibfield  {title} {\bibinfo {title} {Measurement of the
  $6s\ensuremath{-}7p$ transition probabilities in atomic cesium and a revised
  value for the weak charge ${Q}_{W}$},\ }\href
  {https://doi.org/10.1103/PhysRevA.66.020101} {\bibfield  {journal} {\bibinfo
  {journal} {Phys. Rev. A}\ }\textbf {\bibinfo {volume} {66}},\ \bibinfo
  {pages} {020101} (\bibinfo {year} {2002})}\BibitemShut {NoStop}%
\bibitem [{\citenamefont {Dzuba}\ and\ \citenamefont
  {Flambaum}(2000)}]{DzubaBeta}%
  \BibitemOpen
  \bibfield  {author} {\bibinfo {author} {\bibfnamefont {V.~A.}\ \bibnamefont
  {Dzuba}}\ and\ \bibinfo {author} {\bibfnamefont {V.~V.}\ \bibnamefont
  {Flambaum}},\ }\bibfield  {title} {\bibinfo {title} {Off-diagonal hyperfine
  interaction and parity nonconservation in cesium},\ }\href
  {https://doi.org/10.1103/PhysRevA.62.052101} {\bibfield  {journal} {\bibinfo
  {journal} {Phys. Rev. A}\ }\textbf {\bibinfo {volume} {62}},\ \bibinfo
  {pages} {052101} (\bibinfo {year} {2000})}\BibitemShut {NoStop}%
\bibitem [{\citenamefont {Zhang}\ \emph {et~al.}(2026)\citenamefont {Zhang},
  \citenamefont {Feng}, \citenamefont {Gorchtein}, \citenamefont {Jin},
  \citenamefont {Liu},\ and\ \citenamefont {Seng}}]{Zhang:2026utt}%
  \BibitemOpen
  \bibfield  {author} {\bibinfo {author} {\bibfnamefont {Z.-l.}\ \bibnamefont
  {Zhang}}, \bibinfo {author} {\bibfnamefont {X.}~\bibnamefont {Feng}},
  \bibinfo {author} {\bibfnamefont {M.}~\bibnamefont {Gorchtein}}, \bibinfo
  {author} {\bibfnamefont {L.-C.}\ \bibnamefont {Jin}}, \bibinfo {author}
  {\bibfnamefont {C.}~\bibnamefont {Liu}},\ and\ \bibinfo {author}
  {\bibfnamefont {C.-Y.}\ \bibnamefont {Seng}},\ }\href@noop {} {\bibinfo
  {title} {{Lattice QCD determination of the $\gamma Z$ box contribution to the
  proton weak charge}}} (\bibinfo {year} {2026}),\ \Eprint
  {https://arxiv.org/abs/2601.06582} {arXiv:2601.06582 [hep-lat]} \BibitemShut
  {NoStop}%
\bibitem [{\citenamefont {{Jefferson Lab Qweak Collaboration}}(2018)}]{Qweak}%
  \BibitemOpen
  \bibfield  {author} {\bibinfo {author} {\bibnamefont {{Jefferson Lab Qweak
  Collaboration}}},\ }\bibfield  {title} {\bibinfo {title} {{Precision
  measurement of the weak charge of the proton}},\ }\href
  {https://doi.org/10.1038/s41586-018-0096-0} {\bibfield  {journal} {\bibinfo
  {journal} {Nature}\ }\textbf {\bibinfo {volume} {557}},\ \bibinfo {pages}
  {207–211} (\bibinfo {year} {2018})},\ \bibinfo {note} {{Proton weak charge
  $Q_W^p = 0.0719 \pm 0.0045$}}\BibitemShut {NoStop}%
\bibitem [{\citenamefont {{SLAC E158 Collaboration, P. L. Anthony et
  al.}}(2005)}]{E158}%
  \BibitemOpen
  \bibfield  {author} {\bibinfo {author} {\bibnamefont {{SLAC E158
  Collaboration, P. L. Anthony et al.}}},\ }\bibfield  {title} {\bibinfo
  {title} {{Precision Measurement of the Weak Mixing Angle in Møller
  Scattering}},\ }\href {https://doi.org/10.1103/PhysRevLett.95.081601}
  {\bibfield  {journal} {\bibinfo  {journal} {Phys. Rev. Lett.}\ }\textbf
  {\bibinfo {volume} {95}},\ \bibinfo {pages} {081601} (\bibinfo {year}
  {2005})},\ \bibinfo {note} {{Extracted weak charge of the electron $Q_W^e =
  -0.053(11)$}}\BibitemShut {NoStop}%
\bibitem [{\citenamefont {Dzuba}\ \emph {et~al.}(2017)\citenamefont {Dzuba},
  \citenamefont {Flambaum},\ and\ \citenamefont {Stadnik}}]{Dzuba2017PRL}%
  \BibitemOpen
  \bibfield  {author} {\bibinfo {author} {\bibfnamefont {V.~A.}\ \bibnamefont
  {Dzuba}}, \bibinfo {author} {\bibfnamefont {V.~V.}\ \bibnamefont
  {Flambaum}},\ and\ \bibinfo {author} {\bibfnamefont {Y.~V.}\ \bibnamefont
  {Stadnik}},\ }\bibfield  {title} {\bibinfo {title} {{Probing low-mass vector
  bosons with parity nonconservation and nuclear anapole moment measurements in
  atoms and molecules}},\ }\href
  {https://doi.org/10.1103/PhysRevLett.119.223201} {\bibfield  {journal}
  {\bibinfo  {journal} {Phys. Rev. Lett.}\ }\textbf {\bibinfo {volume} {119}},\
  \bibinfo {pages} {223201} (\bibinfo {year} {2017})}\BibitemShut {NoStop}%
\bibitem [{\citenamefont {Marciano}\ and\ \citenamefont
  {Rosner}(1990)}]{MarcianoSTU}%
  \BibitemOpen
  \bibfield  {author} {\bibinfo {author} {\bibfnamefont {W.~J.}\ \bibnamefont
  {Marciano}}\ and\ \bibinfo {author} {\bibfnamefont {J.~L.}\ \bibnamefont
  {Rosner}},\ }\bibfield  {title} {\bibinfo {title} {Atomic parity violation as
  a probe of new physics},\ }\href
  {https://doi.org/10.1103/PhysRevLett.65.2963} {\bibfield  {journal} {\bibinfo
   {journal} {Phys. Rev. Lett.}\ }\textbf {\bibinfo {volume} {65}},\ \bibinfo
  {pages} {2963} (\bibinfo {year} {1990})}\BibitemShut {NoStop}%
\bibitem [{\citenamefont {Gorchtein}\ and\ \citenamefont
  {Spiesberger}(2026)}]{Gorchtein:2026ctl}%
  \BibitemOpen
  \bibfield  {author} {\bibinfo {author} {\bibfnamefont {M.}~\bibnamefont
  {Gorchtein}}\ and\ \bibinfo {author} {\bibfnamefont {H.}~\bibnamefont
  {Spiesberger}},\ }\href@noop {} {\bibinfo {title} {{Parity violation in
  atoms: neutrino-mediated long range forces and finite nuclear size}}}
  (\bibinfo {year} {2026}),\ \Eprint {https://arxiv.org/abs/2605.17059}
  {arXiv:2605.17059 [nucl-th]} \BibitemShut {NoStop}%
\end{thebibliography}
\end{document}